\documentclass[aps,pre,twocolumn,showpacs]{revtex4}
\usepackage{graphicx}
\usepackage{bm}
\usepackage{amsmath}
\begin{document}


\title{Evolving momentum-projected densities in 
billiards \\ with quantum states} 

\author{Debabrata Biswas}
\affiliation{
Theoretical Physics Division, Bhabha Atomic Research Centre, 
Mumbai 400 085, INDIA}

\begin{abstract}
The classical Liouville density on the constant energy surface 
reveals a number of interesting features  when the initial density has
no directional preference. It has been shown (Physical Review Letters, 
{\bf 93}, 204102 (2004)) that the eigenvalues and
eigenfunctions of the momentum-projected density evolution operator
have a correspondence with the quantum Neumann energy eigenstates  in billiard systems. 
While the classical eigenfunctions are well approximated 
by the quantum Neumann eigenfunctions, the classical eigenvalues are of
the form $\{f(\sqrt{E_n} v t)\}$ where $\{E_n\}$ are close to the quantum
Neumann eigenvalues and $v$ is the speed of the classical particle.
Despite the approximate nature of the correspondence, we 
demonstrate here that the exact quantum Neumann eigenstates can be 
used to expand and evolve an  arbitrary classical density on the energy surface
projected on to the configuration space. 
For the rectangular and stadium billiards, 
results are compared with the actual evolution of the density using classical trajectories.

\end{abstract}  
\maketitle


\newcommand{\be}{\begin{equation}}
\newcommand{\ee}{\end{equation}}
\newcommand{\bea}{\begin{eqnarray}}
\newcommand{\eea}{\end{eqnarray}}
\newcommand{\Lop}{{\cal L}}
\newcommand{\Pop}{{\cal P}}
\newcommand{\DB}[1]{\marginpar{\footnotesize DB: #1}}
\newcommand{\q}{\vec{q}}
\newcommand{\kt}{\tilde{k}}
\newcommand{\Lopn}{\tilde{\Lop}}
\newcommand{\noi}{\noindent}

\section{Introduction}
\label{sec:intro}

The phase space density provides an alternate route 
for the study of classical dynamics that is both interesting and
practical. Its evolution is governed by the 
Perron-Frobenius (PF) operator, $\Lop^t$,

\bea
\rho({\bm x},t) = 
\Lop^t \circ \rho({\bm x}) = \int \delta({\bm x} - {\bm f}^t({\bm x}')) 
\rho({\bm x}') d{\bm x'} \label{eq:defPF}
\eea

\noindent
where $\rho({\bm x})$ refers to the phase space density at time $t = 0$, 
${\bm x} = ({\bm q},{\bm p})$ is  a point in phase space 
and ${\bm f}^t({\bm x} )$ is its position at time $t$. 
Equivalently, one may solve the Liouville equation

\be
{\partial \rho \over \partial t} = \{H,\rho\}
\ee

\noi
to determine the evolution of the density. 
A knowledge of the spectral decomposition 
of $\Lop^t$ allows one to evaluate correlations, 
averages and other quantities
of interest \cite{chaosbook,lasota,gaspard}. 

For integrable Hamiltonian dynamics, the eigenvalues and eigenfunctions
of $\Lop^t$ are well known. In terms of the action-angle variables $({\bm I},{\bm \theta})$,
the eigenfunctions are 

\be
\rho_{\bm{I'},\bm{n}}({\bm I},\bm{\theta}) = \frac{1}{{(2\pi)}^{f/2}} 
\delta({\bm I} - {\bm I'})~e^{i {\bm n}.{\bm \theta}} 
\label{eq:int_efunc}
\ee

\noi
where ${\bm I'}$ labels the torus on which the eigenfunction has its support,
$\{{\bm n} | {\bm n} \in {\bm Z}\}$ is a measure of the spatial variation 
of the eigenfunction and $f$ is the number of degrees of freedom.
The corresponding eigenvalues are

\be
\Lambda_{\bm{n}} = e~^{i \bm{n}.{\bm \omega} t} \label{eq:int_eval}
\ee

\noi
where $\bm{\omega}({\bm I}) = \partial H /\partial {\bm I}$ are the frequencies on the
torus. Thus, the spectrum of an integrable system is discrete on a given torus ${\bm I}'$.
Note that $\bm{\omega}$ generally depends on $\bm{I}'$ so that
on the constant energy surface, the spectrum is continuous \cite{gaspard,brumer,narosa98,sridhar06}.

The density, projected on to the configuration space, is the object of 
study in this paper. We shall be interested in the special case where the 
phase space density is constrained to the constant energy surface. Further, we shall 
assume that the initial density has no directional preference 
and its dependence on momentum arises solely through its magnitude 
in the energy conservation requirement.  
Such an initial density can be expressed as~:

\be
\rho_E(\bm{p}, \bm{q}) = \delta(E_0 - H(\bm{p}, \bm{q})) g(\bm{q}).
\ee

\noindent
Note that the hamiltonian nature of the flow ensures that the density
remains confined to the constant energy surface for all times. 
Its projection to the configuration space can be expressed as

\be
\rho_P(\bm{q}) = \int~d\bm{p}~\rho_E(\bm{p}, \bm{q})
\ee

\noindent
where the subscripts $P$ and $E$ denote `momentum projection' and
restriction to the constant energy surface respectively. 

We are interested here in the time evolution of the momentum-projected
density $\rho_P(\bm{q})$. Thus,
we wish to determine the kernel $K_P(\bm{q},\bm{q}',t)$ of the
operator $\Lop_P^t$ which evolves $\rho_P(\bm{q})$:

\be
\Lop_P^t \circ  \rho_P(\bm{q}) = \int d\bm{q}'~K_P(\bm{q},\bm{q}',t) ~\rho_P(\bm{q}').
\ee

\noindent
A straightforward way of arriving at the form of $K_P(\bm{q},\bm{q}',t)$
is by projecting the initial density $\rho_E$ to the configuration space 
after evolving it for a time $t$ i.e.

\bea 
\Lop_P^t \circ \rho_P({\bm q})& = & \int~d{\bm p}~\Lop^t\circ\rho_E({\bm q},
{\bm p}) \nonumber \\
& = & \int d{\bm p} \int d{\bm q}' d{\bm p}'~ \delta({\bm x} - {\bm f}^t({\bm x}')) \nonumber \\
& ~ & ~~~~ \times \delta(E_0 - H(\bm{p}', \bm{q}')) g(\bm{q}')
\eea

\noi
For standard hamiltonians $H({\bm p},{\bm q}) = {\bm p}^2 + V({\bm q})$ (assuming for convenience
$2m = 1$), the projected density 
$\rho_P(\bm{q}) = \int~d{\bm p} \rho_E({\bm p},{\bm q}) = \pi g(\bm{q})$. Thus

\be
\Lop_P^t \circ \rho_P({\bm q}) =   \int d{\bm q}' K_P(\bm{q},\bm{q}',t;E_0) \rho_P({\bm q}')
\label{eq:PF_Projected}
\ee

\noindent
where

\be
K_P(\bm{q},\bm{q}',t; E_0) = {1 \over \pi} \int 
d{\bm p} d{\bm p}' \delta({\bm x} - f^t(\bm{x}')) \delta(E_0 - H({\bm x}'))
\label{eq:Projected_kernel}
\ee 

\noi
and ${\bm x}' = (\bm{p}', \bm{q}')$. Note that the kernel $K_P$ is parameterized 
by $E_0$, the energy at which the initial phase space density is constrained. In 
general, if the starting point is a momentum projected density $\rho_P(\bm{q})$, $\Lop_P^t$ 
evolves $\rho_P(\bm{q})$ isotropically at an energy $E_0$.

We are interested here in the time evolution of such a projected-density
$\rho_P({\bm q})$ using the eigenvalues and eigenfunctions of $\Lop_P^t$ \cite{not_the_same}. 
For billiards \cite{prl2004,pramana2005},       
the eigenfunctions of $\Lop_P^t$  are well approximated 
by the quantum Neumann eigenfunctions, while its eigenvalues are of
the form $\{f(\sqrt{E_n} v t)\}$ where $\{E_n\}$ are close to the quantum
Neumann eigenvalues, $v$ being the speed of the classical particle.
Despite the approximate nature of the correspondence, we shall
demonstrate in this communication that the exact quantum Neumann eigenstates can be 
used to expand and evolve an  arbitrary classical projected-density.

The paper is organized as follows. In section \ref{sec:projected}, we
deal with the projected density evolution operator for billiards
and discuss some of its general properties. In section \ref{sec:estates},
we explicitly show how a projected-density evolves in an 
integrable system and summarize the results \cite{prl2004,pramana2005}
on the eigenvalues and eigenfunctions of $\Lop_P^t$ 
for billiards in general.  
Section \ref{sec:numerics} contains numerical demonstration that quantum 
states can be used to evolve classical projected densities in an
integrable as well as a chaotic system. Finally a summary  and
discussion of the results is provided in the concluding section.

\section{The projected density evolution operator for billiards}
\label{sec:projected}

We shall henceforth restrict ourselves to billiards where the eigenfunctions of 
$\Lop_P^t$ have interesting properties. 
In a classical billiard, a point particle moves freely on a plane inside an 
enclosure and reflects specularly from the boundary.
Depending on the shape of the boundary, billiards exhibit the
entire range of behaviour observed in other dynamical systems.
The quantum billiard problem consists of determining the eigenvalues and
eigenfunctions of the Helmholtz equation

\be 
\nabla^2 \psi({\bm q}) + k^2 \psi({\bm q}) = 0
\ee

\noindent
with $\psi({\bm q}) = 0$ (Dirichlet) or 
$\hat{\bm n}.\nabla\psi = 0$ (Neumann boundary condition; $\hat{\bm n}$
is the unit normal)
on the boundary. Apart from being a paradigm in the field of
classical and quantum chaos, billiards have relevance in a variety
of contexts including the motion of ultra-cold atoms confined by laser beams
(the so called ``optical billiards'') \cite{raisen,davidson}. 
The Helmholtz equation describing the quantum
billiard problem also describes acoustic
waves, modes in microwave cavities and has relevance
in studies on ``quantum wells'', ``quantum corrals'', mesoscopic 
systems and nanostructured materials.
The Neumann boundary condition has important manifestation in
acoustic waves, surface water waves, TE modes in cavities and
modes of a drum with stress-free boundaries \cite{moti} as 
well as excitations of Bose-Einstein condensates in billiards \cite{raisen2004}. 

As the motion inside the enclosure is 
free in a billiard, the magnitude of the momentum is conserved. The kernel for
a billiard is thus

\be
 K_P(\bm{q},\bm{q'},t) =  {1\over \pi} \int~d{\bm p} d{\bm p}'
\delta({\bm q} - {\bm q'}^t)
\delta({\bm p} - {\bm p'}^t) \delta({\bm p}_0^2 - {\bm p}'^2) 
\label{eq:Projected_kernel1}
\ee

\noi
where ${\bm q'}^t = {\bm q'}^t({\bm q}',{\bm p}')$,  ${\bm p'}^t = {\bm p'}^t({\bm q}',{\bm p}')$
is the position after time $t$ of an initial phase space point $({\bm q'},{\bm p'})$.
It is assumed henceforth that the kernel $K_P$ depends on the parameter 
$E_0 = p_0^2$.

It is convenient to treat ${\bm p}$ in polar 
coordinates ($p,\varphi$) where $p$ is the magnitude and 
$\varphi$ is the angle that ${\bm p}$ makes with a given axis.
Transforming from ($p_x,p_y$) to ($p,\varphi$) and
noting that $p$ is conserved, the ${\bm p'}$ 
integration in Eq.~(\ref{eq:Projected_kernel1}) can be simplified as
(see appendix \ref{app}):

\bea
& \; & \int dp_x' dp_y' 
\delta\bigl(p_x - {p_x'}^t({\bm q'},{\bm p'})\bigr) 
\delta\bigl(p_y - {p_y'}^t({\bm q'},{\bm p'})\bigr) 
h({\bm q'},{\bm p'}) \nonumber \\
& ~ & ~~~~=  
\int d\varphi'~\delta(\varphi - \varphi'^t) h({\bm q'},\varphi';p) 
= h({\bm q'},\varphi^*;p)
\eea

\noindent 
where ${p_x'}^t({\bm q'},{\bm p'})$ and ${p_y'}^t({\bm q'},{\bm p'})$ are the
$x$ and $y$ components of the momentum at time $t$ for the initial phase
space coordinate $({\bm q'},{\bm p'})$ and $\varphi^*$ is the value of $\varphi'$ for
which $\varphi'^t({\bm q'},\varphi',p) = \varphi$ \cite{planar}. In the above,

\be
h({\bm q'},\varphi;p) = \delta({\bm q} - {\bm q}'^t({\bm q}',
\varphi,p)) \delta(p^2 - p_0^2).
\ee

\noindent
The kernel thus simplifies as 

\be
K_P(\bm{q},\bm{q'},t)  = {1\over \pi} \int p dp d\varphi~
\delta({\bm q} - {\bm q}'^t({\bm q}';p,\varphi^*)) \delta(p^2 - p_0^2).  \label{eq:kernel_sim}
\ee

\noi
On carrying out the $p$ integration, the kernel can finally be expressed as

\be
K_P(\bm{q},\bm{q'},t; p_0) = {1 \over 2\pi} \int d\varphi~\delta({\bm q} - {\bm q}'^t({\bm q}';p_0,\varphi^*(\varphi))).
 \label{eq:kernel}
\ee

\noi
Thus, the    
time evolution of the projected-density can be determined using Eq.~(\ref{eq:PF_Projected})
and Eq.~(\ref{eq:kernel}).

We list below some properties of the projected-density evolution operator, $\Lop_P^t$:

\begin{enumerate}

\item $\Lop_P^t$ preserves 
positivity of the density. To see this, note that the delta function in 
the kernel ($K_P$) picks up values from the initial density, which, is
positive everywhere by definition. Thus the integrand remains positive
at all times.  

\item{} The operator $\Lop_P^t$ is not multiplicative i.e. $\Lop_P^{t_2+ t_1}
\neq \Lop_P^{t_2} \circ \Lop_P^{t_1}$. A simple way of understanding 
this is by noting that evolution of the projected density is such that
trajectories move out isotropically from each point of the initial 
density. Thus, $\Lop_P^{t_2} \circ (\Lop_P^{t_1} \circ \rho({\bm q}))$ evolves
the initial density isotropically to $\Lop_P^{t_1} \circ \rho({\bm q})$ followed
by an isotropic evolution of this density for
a time $t_2$. This is different from an isotropic evolution of the 
initial density for time $t_1 + t_2$ (see appendix \ref{app2}).

A consequence of nonmultiplicative
evolution is that the eigenvalues of $\Lop_P^t$ are not of the form
$e^{\lambda_i t}$.

\item{} The eigenvalues and eigenfunctions of $\Lop_{P}^t$ have interesting 
properties in billiards \cite{prl2004,pramana2005} as
elaborated in the following section. 
It is found that the eigenfunctions are 
(at least) well approximated by the quantum Neumann eigenfunctions 
while the eigenvalues are  $\{f(\sqrt{E_n} v t)\}$ where 
$\{E_n\}$ are well approximated by the quantum Neumann eigenvalues. 
Here $v$ is the speed 
of the classical particle and $f(x)$ is the asymptotic form of the
Bessel function $J_0(x)$. 

\end{enumerate}
 
\section{Eigenvalues and Eigenfunctions of $\Lop_P^t$}
\label{sec:estates}

\subsection{Integrable Case} 

For integrable systems, the eigenmodes of $\Lop_P^t$ can be determined easily. 
In terms of action-angle variables
$\{ \bm{I,\theta} \}$, an arbitrary initial density $\rho(\bm{I},\bm{\theta})$ 
can be expanded as

\be
\rho(\bm{I},\bm{\theta}) = \sum_{\bm{n}} \int d{\bm I}' C_n(\bm{I}') \rho_{\bm{I}',\bm{n}}(\bm{I},\bm{\theta})
\ee

\noi
where $\rho_{\bm{I}',\bm{n}}$ are the eigenfunctions of $\Lop^t$ (Eq.~\ref{eq:int_efunc})
and the coefficients

\bea
C_{\bm{n}}(\bm{ I}') & = & \int~d\bm{\theta_0}~d\bm{I_0} 
\rho(\bm{I_0},\bm{\theta_0}) \rho^*_{\bm{I'},\bm{n}}(\bm{I_0},\bm{\theta_0}) \nonumber \\
& =  & \frac{1}{{(2\pi)}^{f/2}} \int d{\bm \theta_0}~\rho({\bm I}',\theta_0) e^{-i {\bm n}.{\bm \theta_0}}. \label{eq:int_coeff}
\eea

The evolution of $\rho({\bm I},{\bm \theta})$ in terms of the eigenvalues 
and eigenfunctions of $\Lop^t$ can thus be expressed as

\be
\Lop^t \circ \rho(\bm{I,\theta}) = \sum_{\bm{n}} \int d{\bm I}' C_{\bm{n}}({\bm I}')~
\rho_{\bm{I'},\bm{n}}(\bm{I},\bm{\theta})~e^{i \bm{n}.\bm{\omega(I')}t} \label{eq:expansion}
\ee

Using  Eq.~(\ref{eq:int_efunc}) for the eigenfunctions, one
can express Eq.~(\ref{eq:expansion}) as 

\bea
\Lop^t \circ \rho(\bm{I,\theta}) & = & \sum_{\bm{n}} \int d{\bm I}' C_{\bm{n}}({\bm I}')
\delta({\bm I} - {\bm I}') \tilde{\rho}_{\bm n}
e^{i \bm{n}.\bm{\omega(I')}t} \nonumber \\
& = &  \sum_{\bm{n}} C_{\bm{n}}({\bm I}) e^{i \bm{n}.\bm{\omega(I)}t} 
\tilde{\rho}_{\bm n}({\bm \theta}) \label{eq:int_expand}
\eea

\noi
where 

\be
\tilde{\rho}_{\bm n}({\bm \theta}) = 
\int d{\bm I}~ \rho_{{\bm I'},{\bm n}} = \frac{e^{i{\bm n}.{\bm \theta}}}{{(2\pi)}^{f/2}}.
\ee

Our interest here is in initial densities that are constrained to the
constant energy  surface having the form

\be
\rho_E(\bm{I},\bm{\theta}) = \delta(E -  H(\bm{I})) g(\bm{\theta}). \label{eq:int_density}
\ee

\noi
Thus using Eqns.~(\ref{eq:int_coeff}), (\ref{eq:int_expand}) and (\ref{eq:int_density}) 

\bea
\Lop^t \circ \rho_E(\bm{I,\theta}) & = & \sum_{\bm{n}} \left( \int d{\bm \theta_0}~g({\bm \theta_0}) 
\tilde{\rho}_{\bm n}^*({\bm \theta_0}) \right) \nonumber \\ 
& \times & \delta(E - H({\bm I})) e^{i \bm{n}.\bm{\omega(I)}t} \tilde{\rho}_{\bm n}({\bm \theta})
\eea

\noi
Finally, in order to project  $\rho_E$ to ${\bm \theta}$ space, we must integrate 
$\Lop^t \circ \rho_E(\bm{I,\theta})$ over ${\bm I}$ to obtain 

\bea
\Lop_P^t \circ \rho_P({\bm \theta}) & = & \sum_{\bm{n}} \left(  \int d{\bm \theta_0}~
\rho_P({\bm \theta_0}) \tilde{\rho}_{\bm n}^*({\bm \theta_0})
 \right) \tilde{\rho}_{\bm n}({\bm \theta}) \nonumber \\ 
& \times & \left[ {1 \over \mu} \int d{\bm I} \delta(E - H({\bm I})) 
e^{i \bm{n}.\bm{\omega(I)}t} \right] \label{eq:exp_int} \\
& = & \sum_n \tilde{C}_{\bm n}~ \Lambda_{\bm n}^P ~\tilde{\rho}_{\bm n}({\theta}) 
\label{eq:int_final}
\eea

\noi
where $\rho_P({\bm \theta}) = \int d{\bm I}~ \rho_E({\bm I},{\bm \theta}) = \mu g({\bm \theta})$
and the quantities within (\ldots) and [\ldots] are
respectively the coefficient of expansion ($\tilde{C}_{\bm n}$) of the projected density $\rho_P({\bm \theta})$ 
in terms of  $\tilde{\rho}_{\bm n}$ {\em and}
the corresponding eigenvalue ($\Lambda_{\bm n}^P$) of the projected evolution operator $\Lop_P^t$.    

To illustrate this, we shall consider
the rectangular billiard which is an integrable system. 
The hamiltonian expressed in terms of the actions,
${I_1,I_2}$ is $H(I_1,I_2) = \pi^2(I_1^2/L_1^2 + I_2^2/L_2^2)$
where $L_1,L_2$ are the lengths of the two sides and the mass $m = 1/2$.
With $I_1 = \sqrt{E}L_1\cos(\varphi )/\pi$ and
$I_2 = \sqrt{E} L_2\sin(\varphi )/\pi$, at a given energy, $E$, 
each torus is parameterised by a
particular value of $\varphi$. Transforming from $I_1,I_2$ to $E,\varphi$,
we have 

\bea
\rho_P & = & g({\bm \theta}) {L_1 L_2 \over 2\pi^2} \int dE~ d\varphi~ \delta(E_0 - E) \nonumber \\
& = & {L_1 L_2 \over \pi} g({\bm \theta}) = \mu  g({\bm \theta}).
\eea

\noi
The eigenvalues of $\Lop^t_P$ (denoted as $\Lambda_{\bm n}^P$) are

\bea
\Lambda_{\bm n}^P & =  & {L_1 L_2 \over 2\pi^2 \mu} \int dE d\varphi \delta(E_0 - E) e^{2\pi i t \sqrt{E} [{n_1 \over L_1} \cos\varphi +
{n_2 \over L_2} \sin\varphi]} \nonumber \\
& = & {1 \over 2\pi} \int d\varphi~ e^{i 2\pi l \sqrt{E_{\bm n}} \cos(\varphi - \mu_{\bm n})} \nonumber \\
& =  &  J_0(\sqrt{E_{\bm n}}l)
\eea

\noindent
where $E_n = \pi^2(n_1^2/L_1^2 + n_2^2/L_2^2)$, $\tan \mu_{\bm n} = n_2/n_1$, 
$l = tv$ and 
$J_0$ is the zeroth order Bessel function.
Thus, from Eq.~\ref{eq:int_final}, it follows on putting 
$\rho_P({\bm \theta}) = \tilde{\rho}_{\bm m}({\bm \theta})$  that

\be
\Lop^t_P \circ \tilde{\rho}_{\bm m}({\bm \theta}) =  J_0(\sqrt{E_{\bm m}}l)~ 
 \tilde{\rho}_{\bm m}({\bm \theta})
\ee

\noi
as $\tilde{C}_{\bm n} = \delta_{{\bm m},{\bm n}}$.
Note that ${\bm m} = (m_1,m_2)$ are natural numbers so that 
$e^{i(m_1\theta_1 + m_2\theta_2)}$ are also the 
quantum Neumann eigenfunctions in $\theta$ space while $E_{\bm{m}}$ are
the quantum Neumann energy eigenvalues. 

We end this sub-section with an observation. For the integrable case, the density

\be
\rho_{E,{\bm n}} = {1 \over \mu} \delta(E_0 - H({\bm I})) \frac{e^{i{\bm n}.{\bm \theta}}}{(2\pi)^{f/2}}
\ee 

\noi
on the constant energy surface is an eigenfunction of $\Lop^t$ as 
$\Lop^t \circ \rho_{E,{\bm n}} = e^{i{\bm \omega}.{\bm n}t} \rho_{E,{\bm n}}$. 
Since
 
\be
\int d{\bm I}~\delta(E_0 - H({\bm I})) = \mu,
\ee

\noi
the momentum projection of $\rho_{E,{\bm n}}$ is

\be
\int~d{\bm I}~ \rho_{E,{\bm n}} = \frac{e^{i{\bm n}.{\bm \theta}}}{(2\pi)^{f/2}} = 
\tilde{\rho}_{\bm n}({\bm \theta}).
\ee

\noi
Note that $\tilde{\rho}_{\bm n}({\bm \theta})$ is an eigenfunction of $\Lop_P^t$ i.e.

\be
\Lop_P^t \circ \tilde{\rho}_{\bm n}({\bm \theta}) = \int d{\bm I}~ \Lop^t \circ \rho_{E,{\bm n}} = 
\Lambda_{{\bm n}}^P ~\tilde{\rho}_{\bm n}({\bm \theta})
\ee

\noi
In contrast, for an eigenfunction on the torus (Eq.~\ref{eq:int_efunc}), momentum 
projection yields

\bea
\int d{\bm I}~ \Lop^t \circ \rho_{{\bm I'},{\bm n}} & = & 
\tilde{\rho}_{\bm n}({\bm \theta}) \int d{\bm I} \delta({\bm I} - {\bm I'}) e^{i{\bm n}.{\bm \omega}({\bm I })t} \nonumber \\
& = & e^{i{\bm n}.{\bm \omega}({\bm I'})t} \tilde{\rho}_{\bm n}({\bm \theta})  
\neq \Lambda_{\bm n}^P e^{i\omega({\bm I'})t}
\eea

\noi
Thus the properties of the
projected operator depend on the function space to which the phase space density belongs.

\subsection{Billiards - The Quantum Connection}

It is instructive to see that the quantum Neumann eigenfunctions in
$\bm{q}$ space are also eigenfunctions of $\Lop_P^t$. 
As an illustrative example, 
consider a particle in a one-dimensional box with walls placed at $q = 0$ and
$q = L$. To find the evolution
of the quantum Neumann eigenfunction  
$\psi_n(q) = e^{i k_n q} + e^{-i k_n q}$, 
$k_n = n\pi/L$ under $\Lop_P^t$,
note that in this 1-dimensional case there are only two possible 
values of $\varphi$ i.e. $\varphi = 0, \pi$. Thus

\bea
\Lop^t_P & = & {1\over 2} [ \Lop^t(\varphi=0) + \Lop^t(\varphi=\pi) ] 
\nonumber \\
         & = & {1\over 2} [ \Lop^t(+) +\Lop^t(-) ]
\eea  

\noindent
where $\Lop^t(\pm)$ refer to positive ($\varphi = 0$) 
and negative ($\varphi = \pi$) velocity. The
classical evolution for positive velocity, 
$\Lop^t(+) \circ \psi_n(q)$, is given by

\bea
\left ( e^{i k_n q^{-t}(+v)} + 
e^{-i k_n q^{-t}(+v)}
\right ) 
\eea

\noindent
where $q^{-t}(+v)$ is the position at time $-t$ with initial position $q$
and initial velocity $+v$. Similarly, $\Lop^t(-) \circ \psi_n(q)$ is

\bea
\left ( e^{i k_n (q^{-t}(-v))} + 
e^{-i k_n(q^{-t}(-v))}
\right ) 
\eea

\noindent
with the $-$ sign in $\Lop^t(-)$ denoting negative velocity.
Note that the flow is such that the velocity changes sign at every 
reflection from the walls placed
at $q=0$ and $q=L$. For the flow $q^{-t}(+v)$, the
reflections occur at $t_n^+ = (q + nL)/v$ so that for $t_0^+ < t < t_1^+$,
$q^{-t}(+v) = v(t-t_0^+) = vt - q$. Similarly, for the flow $q^{-t}(-v)$,
the reflections occur at $t_n^- = (L-q + nL)/v$ and for 
$t_0^+ < t < t_1^+$, $q^{-t}(-v) = L - v(t-t_0^-) = 2L - vt - q$.
It follows hence that

\bea
\Lop^t(\pm)\circ \psi_n(q) = e^{i k_n(-q~\pm~vt)} + e^{-i k_n(-q~\pm~vt)}
\eea

\noindent
for all $t$. Thus

\bea
 \Lop^t_P \circ \psi_n(q) =  \cos(k_n v t) \psi_n(q).
\eea

\noindent
In other words, the quantum eigenfunction is also an eigenfunction 
of the projected classical evolution operator $\Lop^t_P$.

For general 2-dimensional billiards, there are two approaches
for determining the eigenmodes of $\Lop_P^t$ \cite{prl2004,pramana2005}. 
Both rely on polygonalization of the billiard boundary \cite{ford,poly}
and the argument that
as the number of sides of the polygon is increased, the modes of the 
smooth billiard are approximated better. In the first approach \cite{arb2}, the 
trace of the $\Lop^t_P$ is related to the trace of the energy dependent
quantum Neumann propagator while in the second, a plane wave expansion 
enables one to conclude that there exists a correspondence between the
quantum Neumann eigenmodes and the modes of $\Lop_P^t$ \cite{arb3,pramana2005}.
We merely state here the
result and refer the interested reader to \cite{prl2004,pramana2005}. For 
$t > 0$,

\bea
\Lop^t_P \circ \psi_n({\bm q}) 
 =    f(k_nvt)~\psi_n({\bm q}) \label{eq:central}
\eea

\noi
where $E_n = k_n^2$ are well approximated by 
the quantum Neumann energy  eigenvalues, 
$f(x) = \sqrt{2/\pi x} \cos(x - \pi/4)$ is
the asymptotic form of $J_0(x)$ and $\psi_n(q)$
are well approximated by the quantum Neumann eigenfunctions. 
For some integrable systems, it is possible to show that
$\{E_n\}$ and $\{\psi_n\}$ are the exact quantum Neumann eigenvalues and
eigenfunctions.

\section{Evolving Classical Densities with Quantum States: Numerical Evidence}
\label{sec:numerics}

As stated earlier, the purpose of this communication is to demonstrate that quantum Neumann
energy eigenfunctions $\{\psi_{\bm n}({\bm q})\}$ 
can be used to evolve an arbitrary classical momentum-projected density
$\rho_P(\bm{q})$, despite the fact that Eq.~(\ref{eq:central}) is generally approximate.

The completeness of the quantum Neumann eigenfunctions allows us to expand
$\rho_P(\bm{q})$ as:

\be
\rho_P(\bm{q}) = \sum_{\bm{n}} C_{\bm{n}} \psi_{\bm{n}}(\bm{q})
\ee

\noi
where 

\be
C_{\bm{n}} = \int~d{\bm q}~\rho_P(\bm{q}) \psi_{\bm{n}}^*(\bm{q}).
\ee

\noi
In the above, it is assumed that 
$\int ~d{\bm q}~\psi_{\bm{n}}^*(\bm{q})\psi_{\bm{m}}(\bm{q}) = 
\delta_{{\bm m},{\bm n}}$.
On using Eq.~(\ref{eq:central}), we have

\be
\Lop_P^t \circ \rho_P(\bm{q}) = \sum_{\bm{n}} C_{\bm{n}}  J_0(\sqrt{E_{\bm{n}}}vt) \psi_{\bm{n}}(\bm{q}).
\label{eq:evolve}
\ee

\noi 
In order to account for the behaviour at small $t$, we use the Bessel function $J_0$
since $f(x)$ is an asymptotic form of $J_0(x)$.  
Note that the evolution of the projected-density, $\rho_P(\bm{q})$, under Eq.~(\ref{eq:evolve})  
should reflect isotropy in momentum.

To demonstrate that Eq.~(\ref{eq:evolve}) indeed gives us the correct classical evolution,
we choose points (initial conditions) in $\bm{q}$-space distributed according 
to the initial density $\rho_P(\bm{q})$ and evolve them in time using classical 
equations of motion. The evolution of these points give us the density $\rho_P(\bm{q},t)$
at any time $t$. This is compared with the density obtained from Eq.~(\ref{eq:evolve}).

In both the examples considered below, we shall consider a hat function as the 
initial density for convenience. The function takes the value unity 
within a square strip of side $\Delta L$ centred at a point $\bm{q}$ and is zero
outside. Thus,  all initial conditions lie uniformly within
the square strip.  
For convenience, we choose the velocity $v = 1$.

\begin{figure*}[tbp]
\begin{center}
\includegraphics[width=4.cm,angle=270]{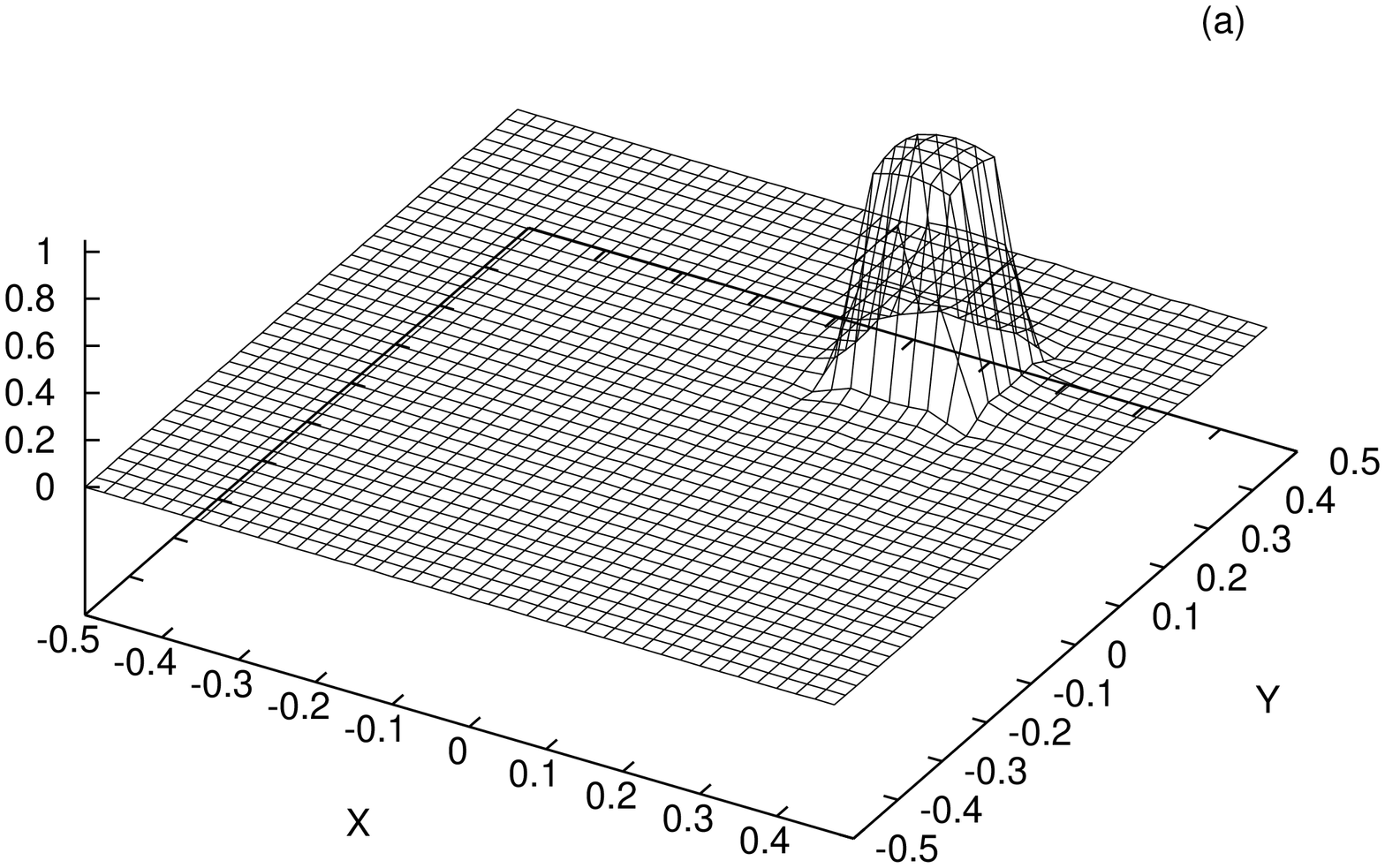}
\includegraphics[width=4.cm,angle=270]{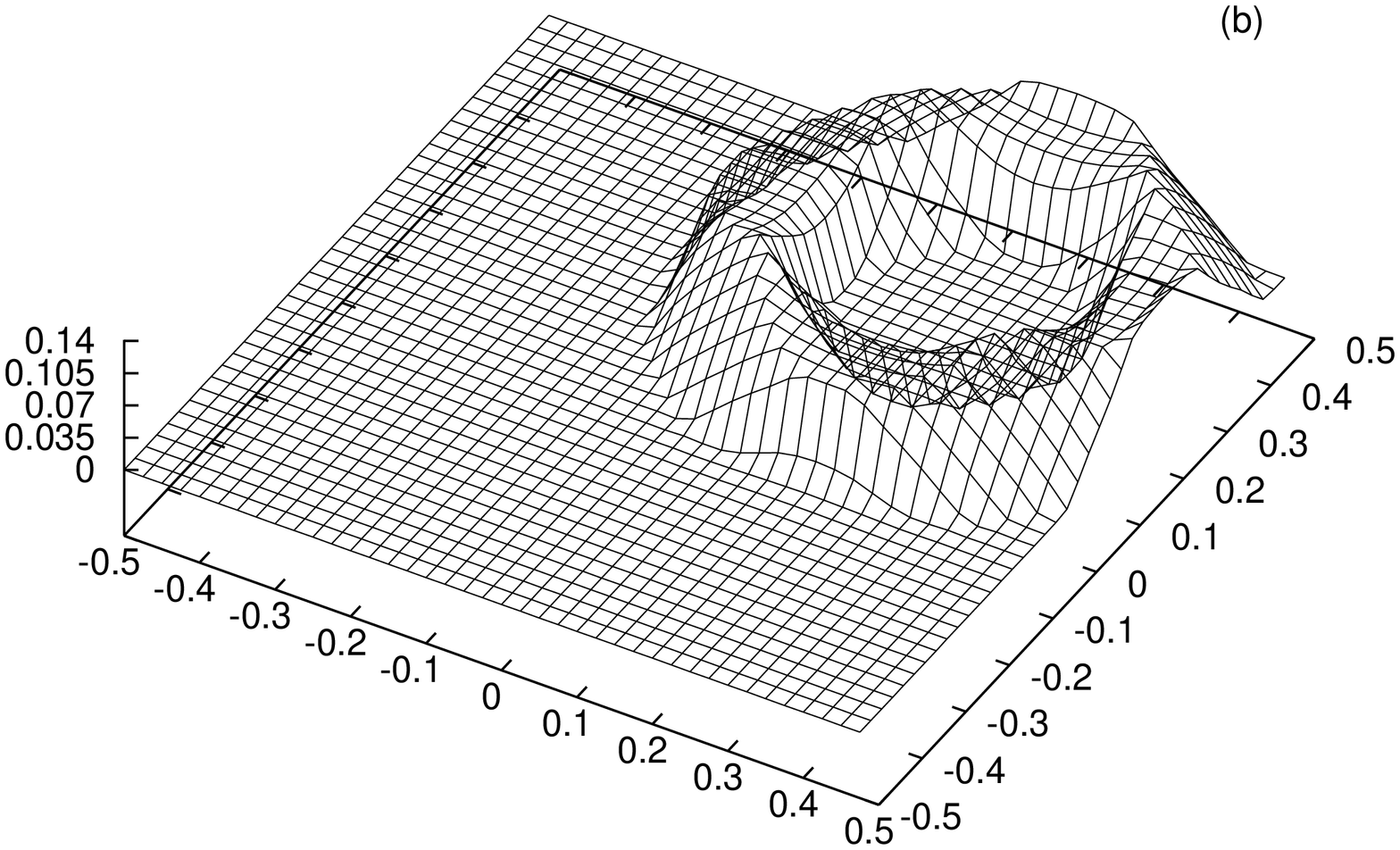}
\includegraphics[width=4.cm,angle=270]{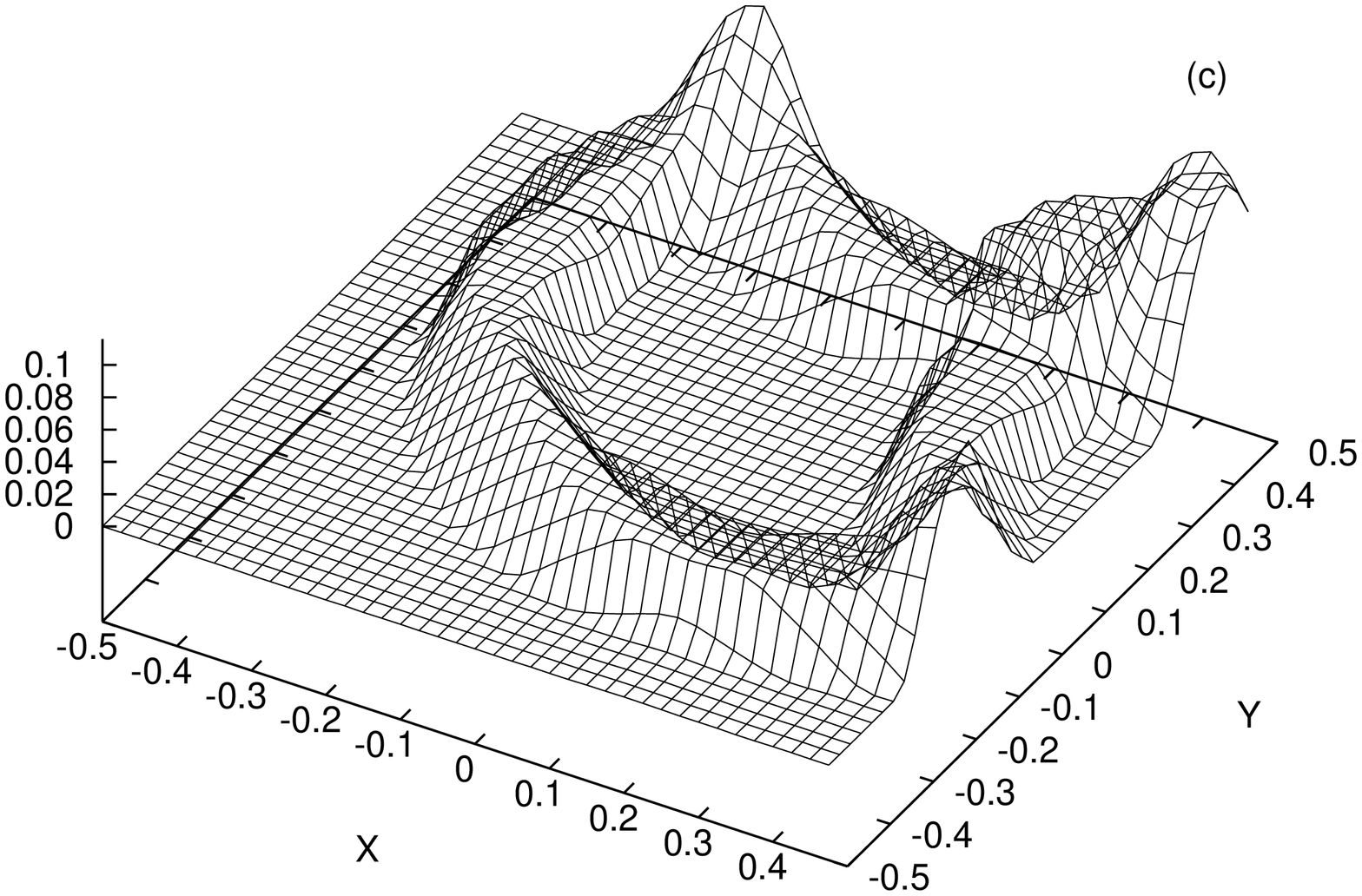}
\includegraphics[width=4.cm,angle=270]{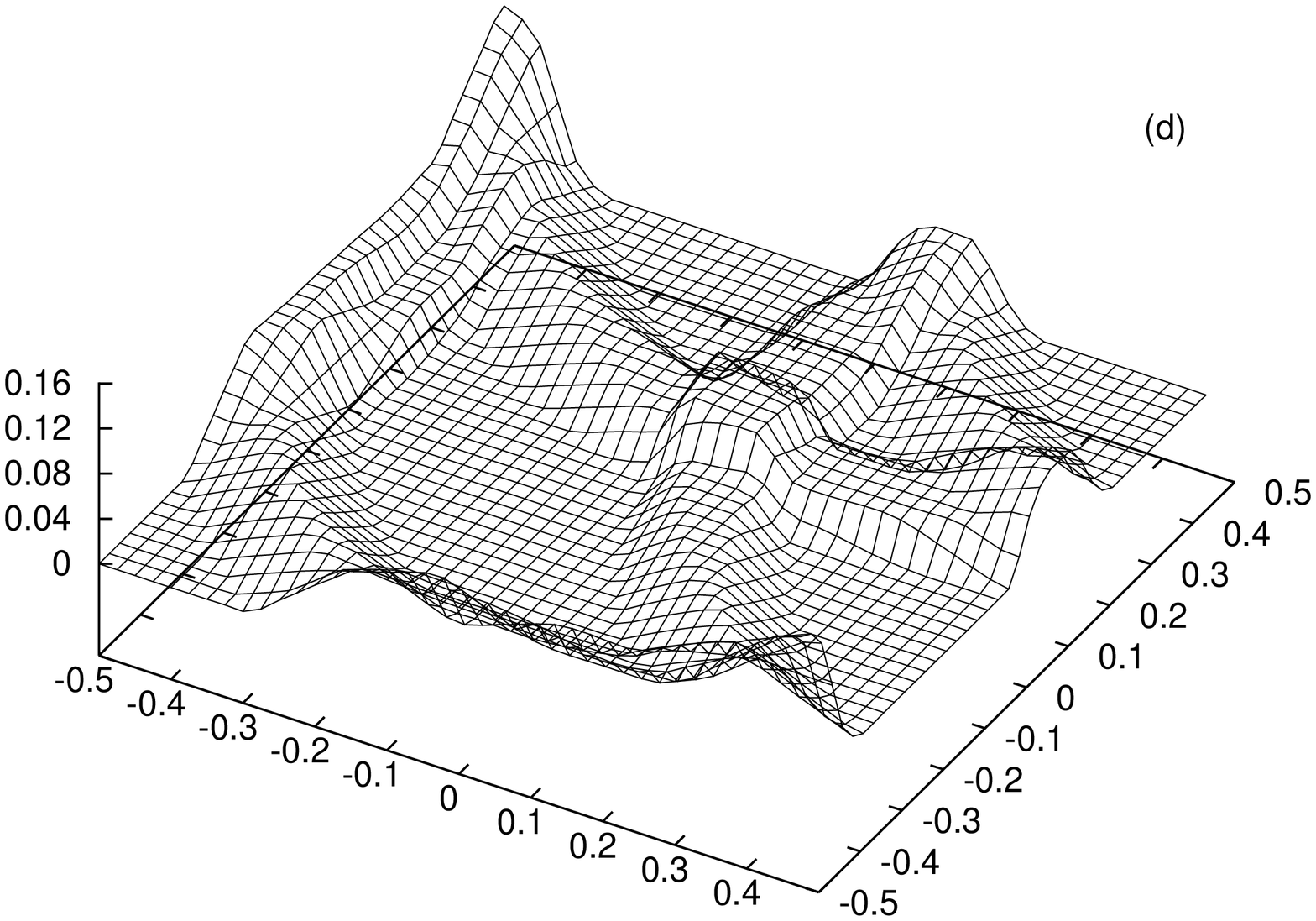}
\includegraphics[width=4.cm,angle=270]{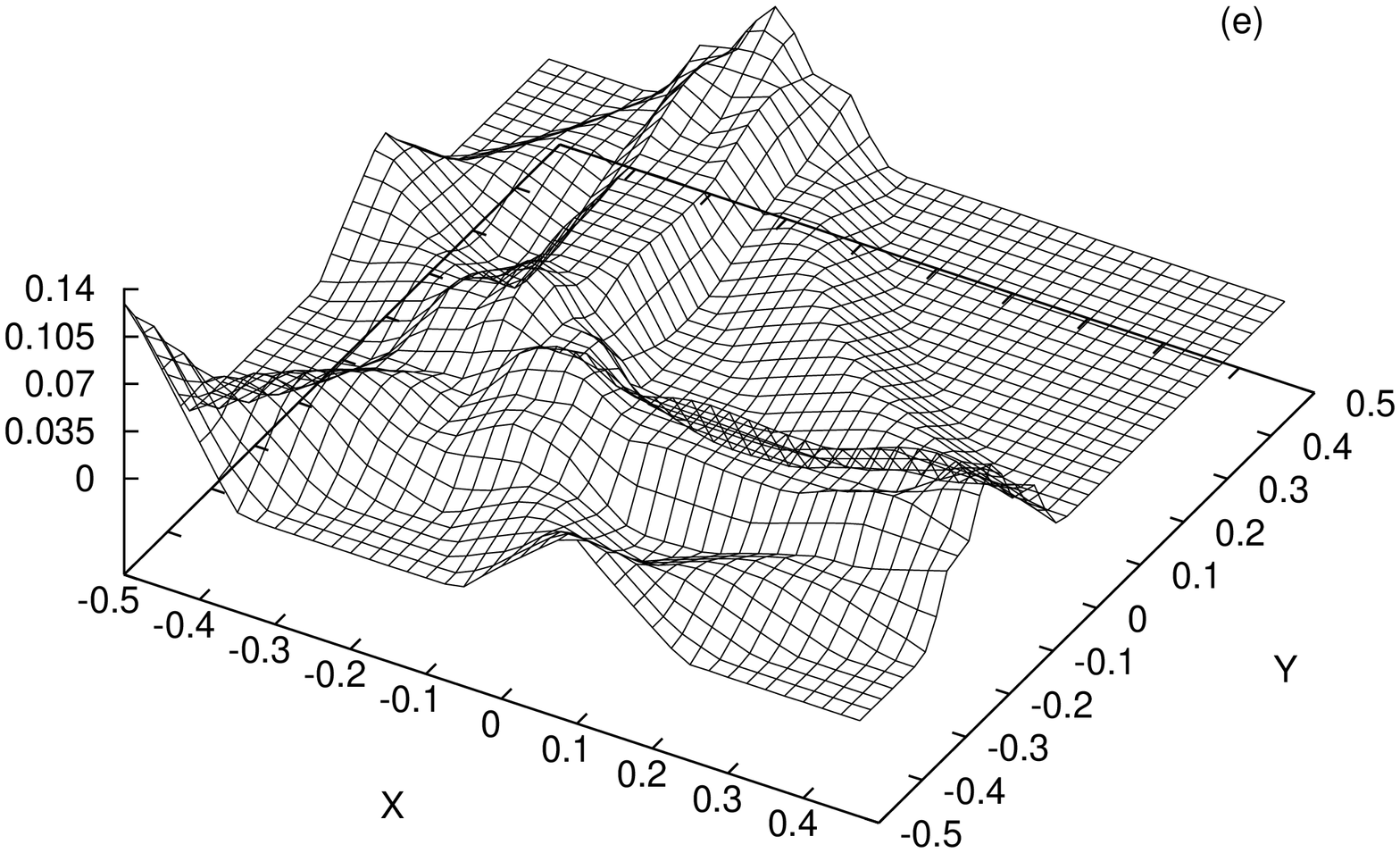}
\end{center}
\caption[ty] {The evolution of a projected-density, $\rho({\bm q})$, in the rectangular billiard
is computed using the exact quantum Neumann eigenfunctions as the eigenfunctions of the projected
Perron-Frobenius operator, $\Lop_P^t$. The initial density is shown in (a) while (b) to (e) 
show the density at t=0.5, 1.0, 1.5, 2.0 respectively. The evolution reflects isotropy in
momentum.}
\label{fig:rect_quant}
\end{figure*}

\begin{figure*}[tbp]
\begin{center}
\includegraphics[width=4.cm,angle=270]{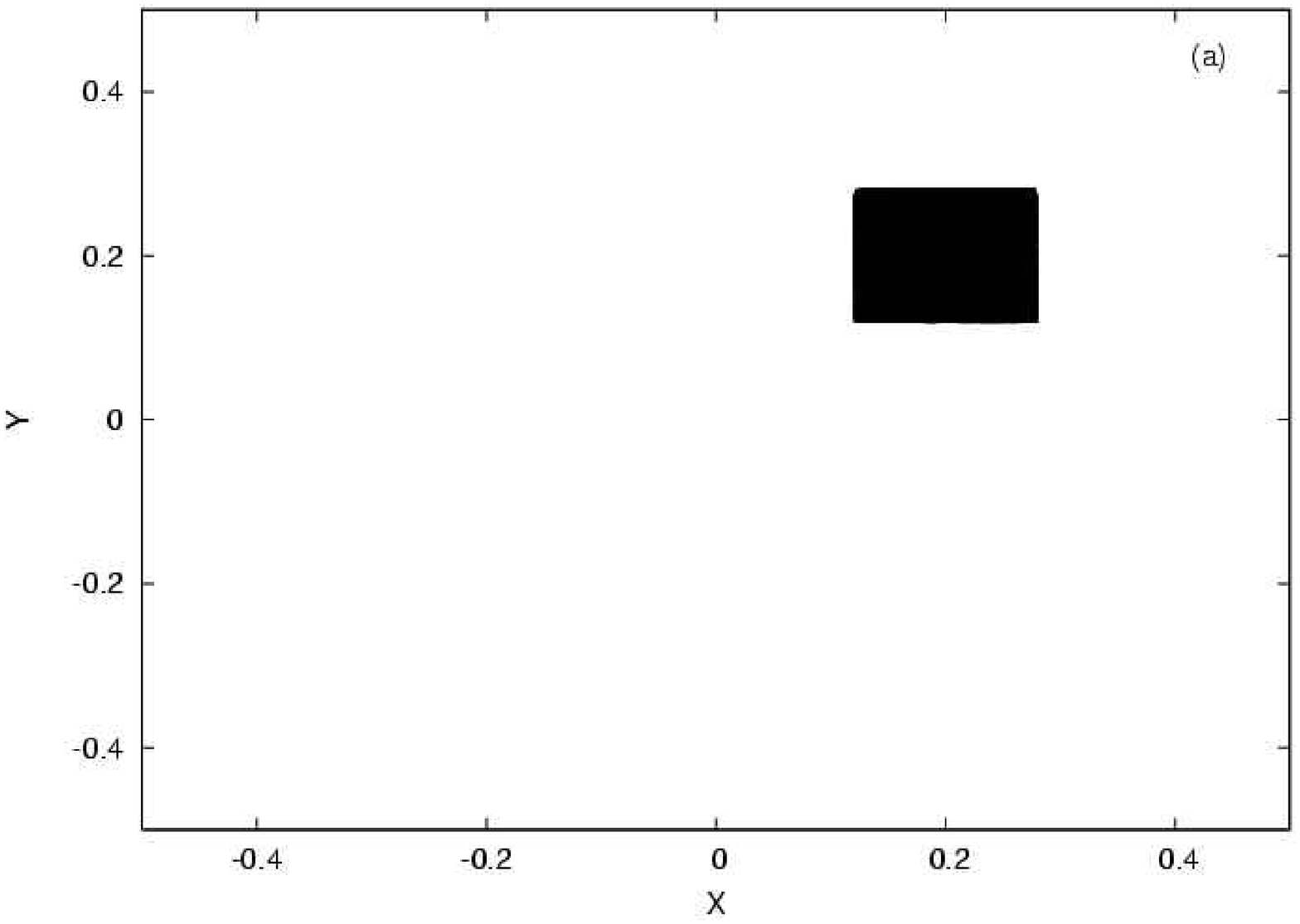}
\includegraphics[width=4.cm,angle=270]{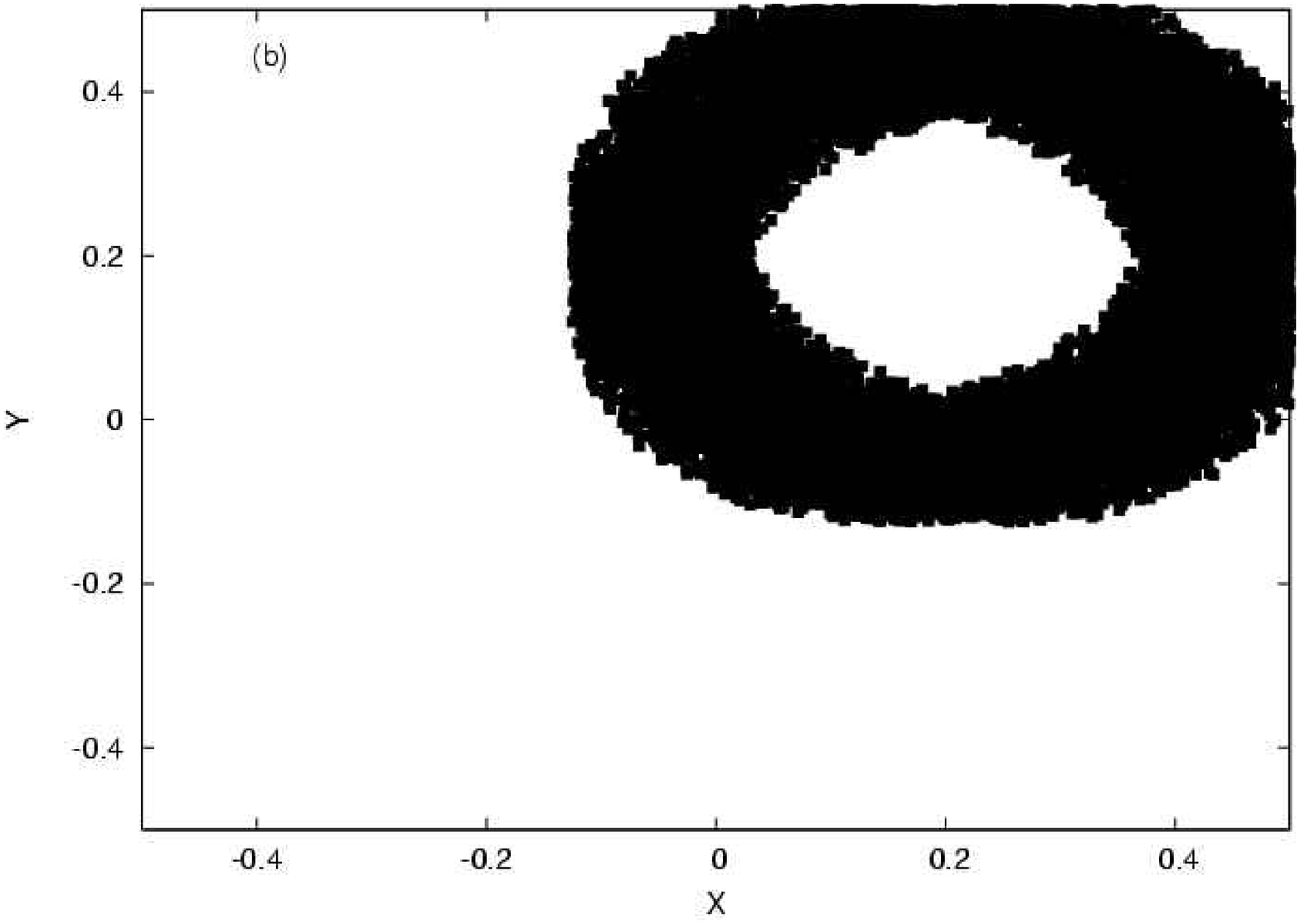}
\includegraphics[width=4.cm,angle=270]{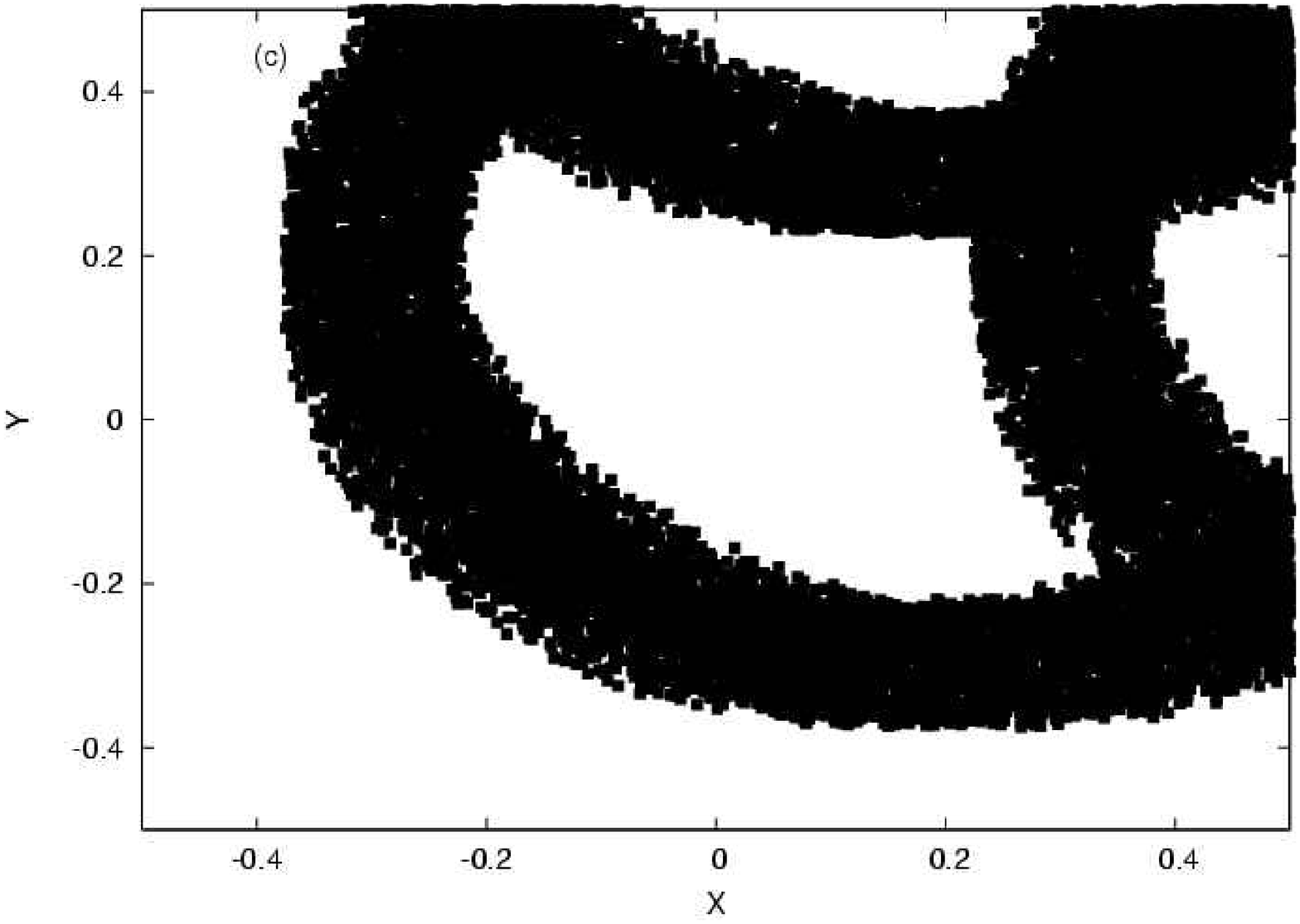}
\includegraphics[width=4.cm,angle=270]{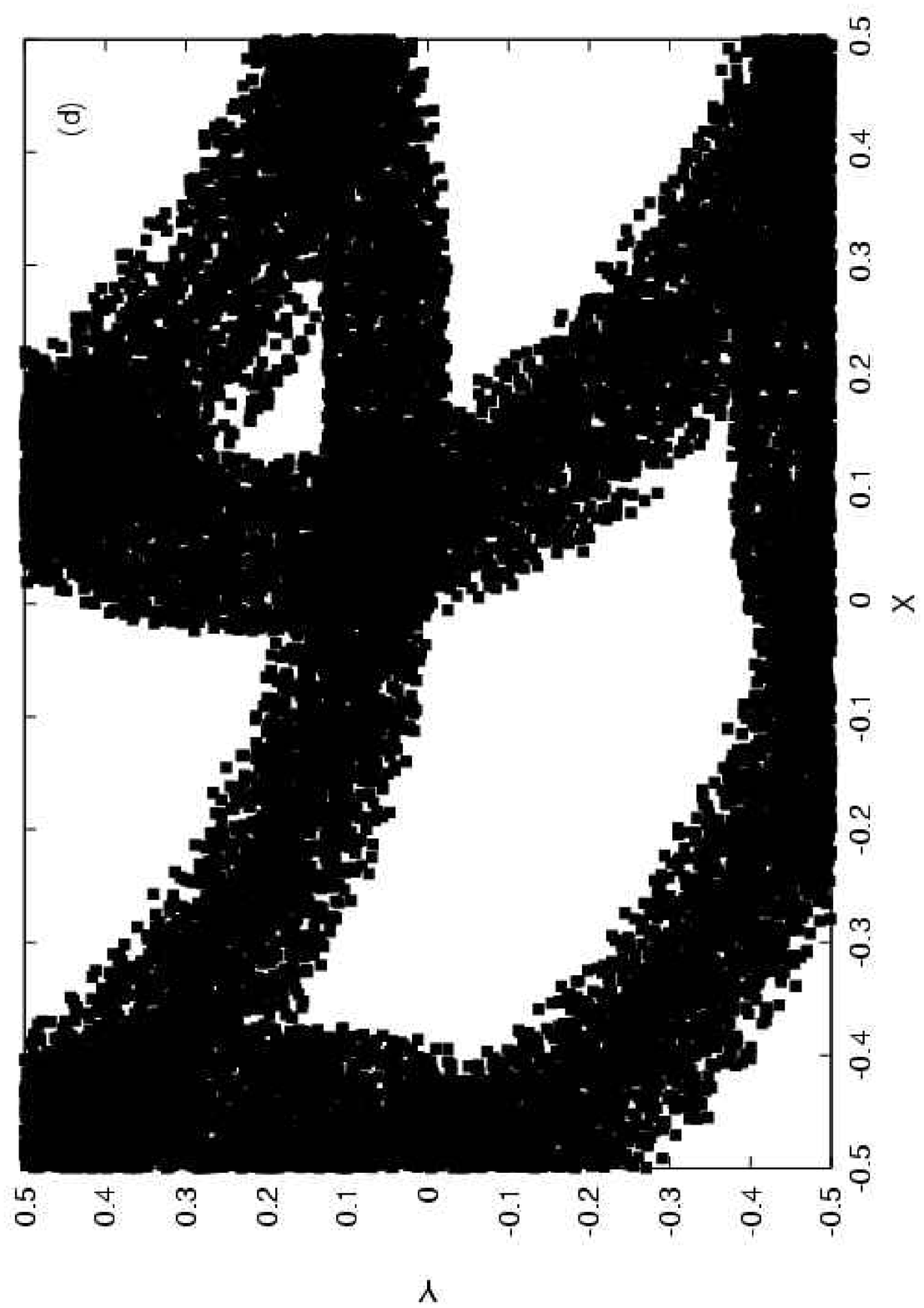}
\includegraphics[width=4.cm,angle=270]{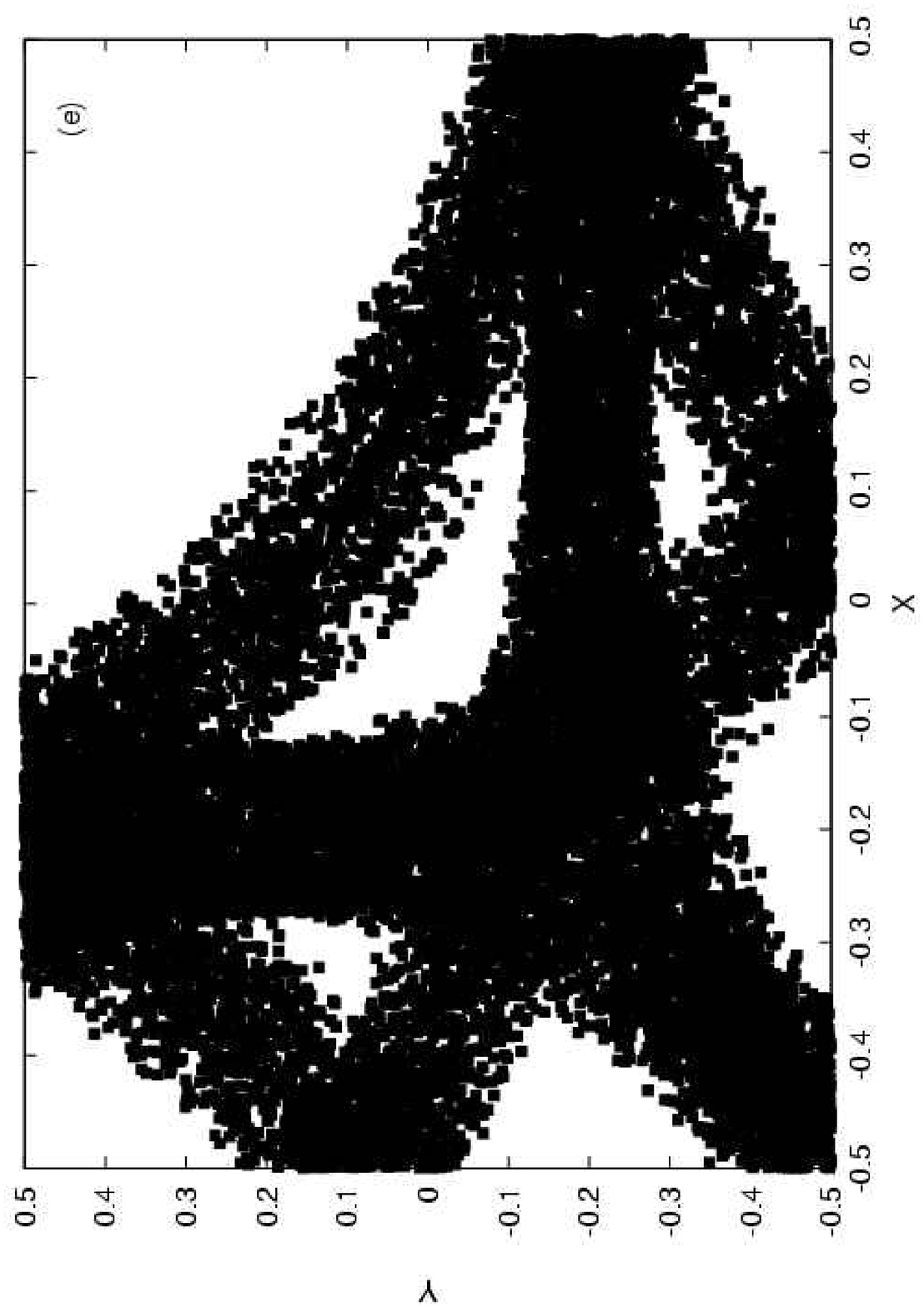}
\end{center}
\caption[ty] {The initial points chosen according to the density in Fig.~1a is shown in
Fig.~2a. These are evolved classically using momentum direction ($\varphi$) 
 distributed uniformly in ($0,2\pi$). The distribution of points at time (b) 0.5 (c) 1.0 (d) 1.5
and (e) 2.0 can be compared with the densities in Fig.~1b to 1e respectively.
}
\label{fig:rect_classical}
\end{figure*}

As a first example, consider a particle in a rectangular box of side $L_1 = L_2 = 1.0$.
Fig.~(\ref{fig:rect_quant}a) shows the initial density centred at (0.2,0.2) with $\Delta L = 0.15$.
The first 5000 quantum Neumann states have been used to 
expand the initial density. 
Figs.~(\ref{fig:rect_quant}b) to (\ref{fig:rect_quant}e) are
the densities at time $t = 0.25, 0.5, 0.75, 1.0$ respectively as computed 
using Eq.~(\ref{eq:evolve}). For comparison, Fig.~\ref{fig:rect_classical} 
is a 2-dimensional view of the density
evolved using classical trajectories  with their momentum distributed 
isotropically \cite{fnote_momentum}.
At time $t = 0$, the points lie in a square patch 
as shown in Fig.~(\ref{fig:rect_classical}a) while  
Figs.~(\ref{fig:rect_classical}b) to (\ref{fig:rect_classical}e) show the points 
as they evolve from Fig.~(\ref{fig:rect_classical}a) at times
$t = 0.25, 0.5, 0.75, 1.0$ respectively. Note that the evolution in
Fig.~1 reflects isotropy in momentum as expected.

\begin{figure*}[tbp]
\begin{center}
\includegraphics[width=4.cm,angle=270]{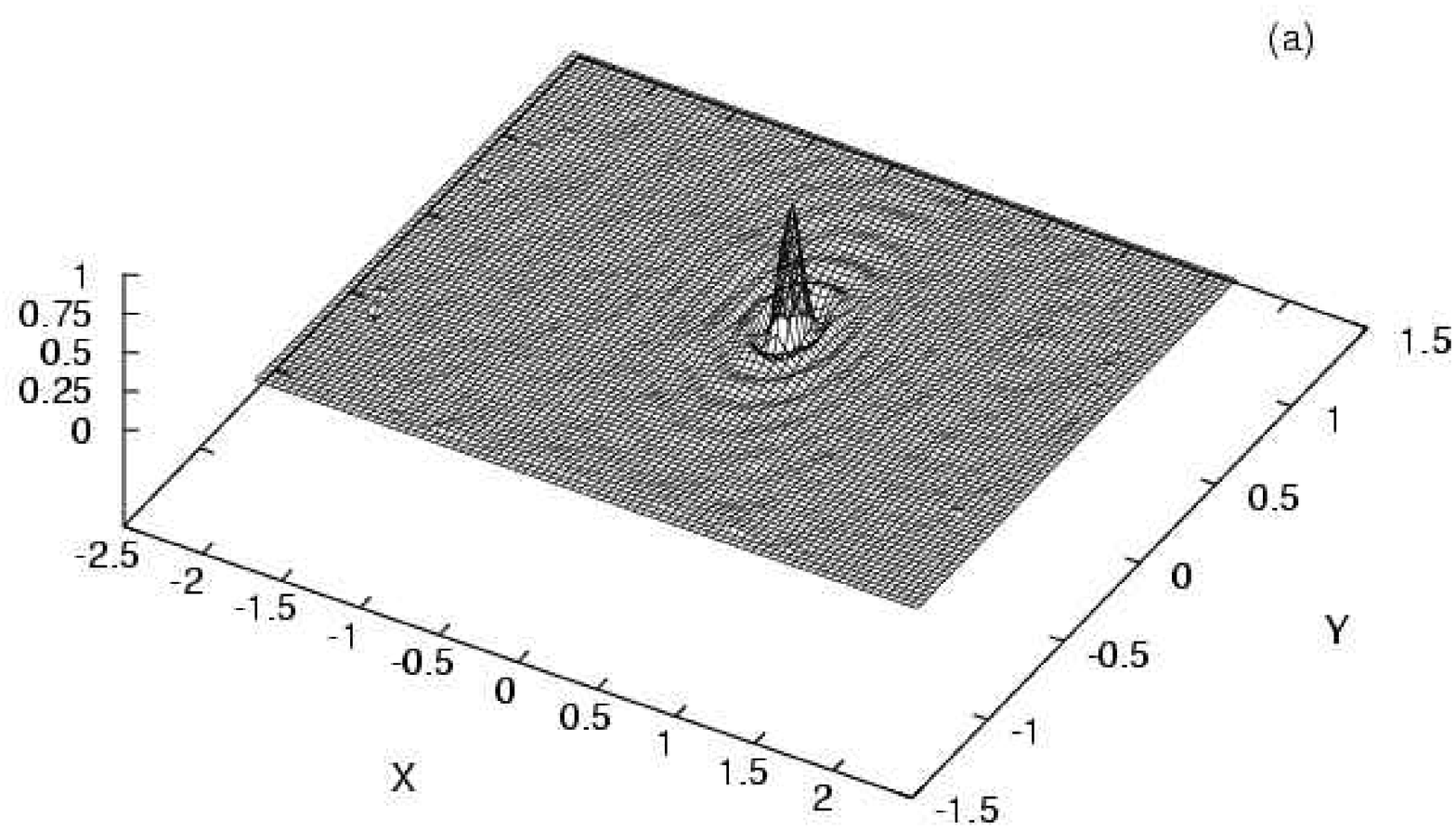}
\includegraphics[width=4.cm,angle=270]{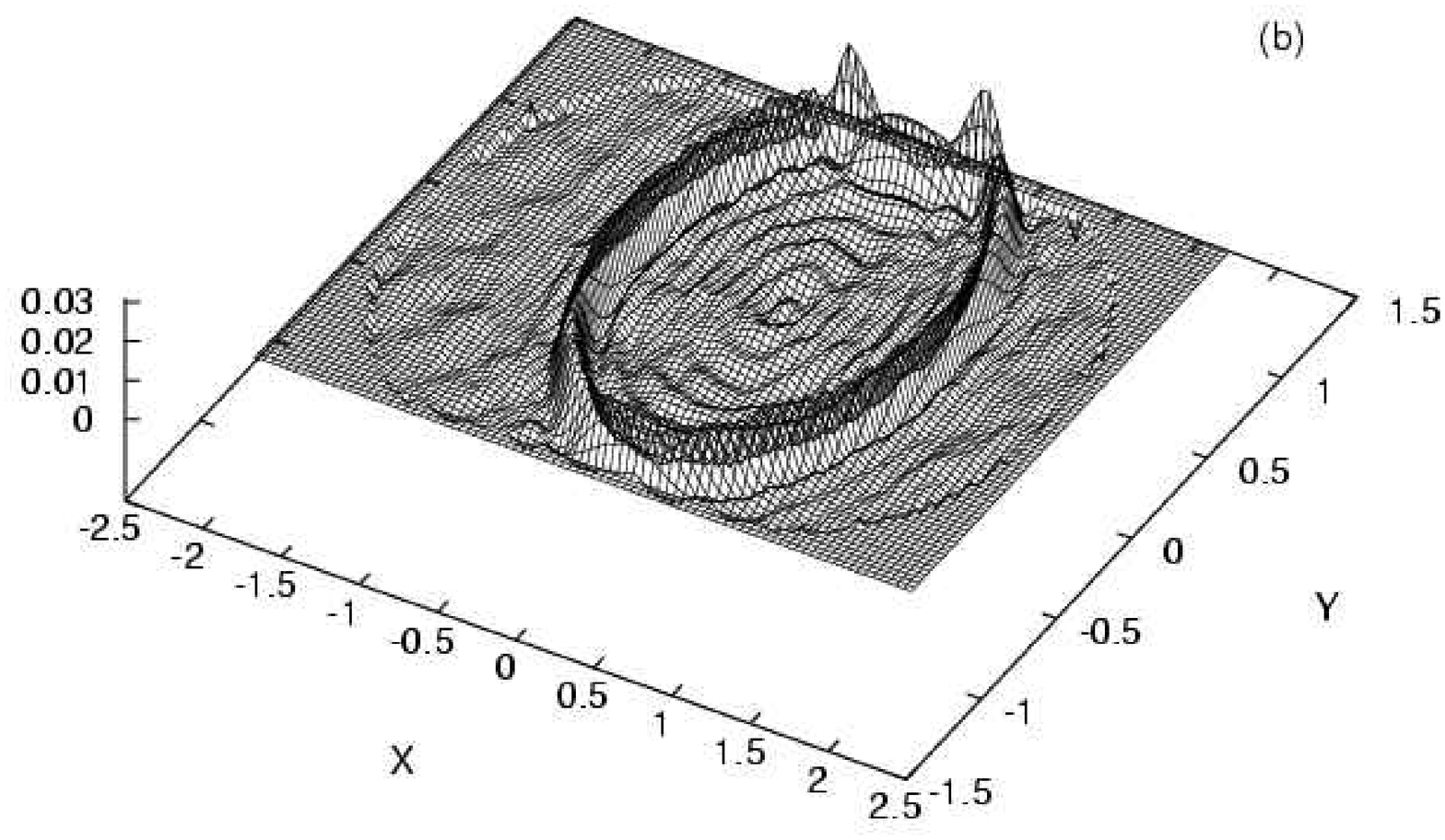}
\includegraphics[width=4.cm,angle=270]{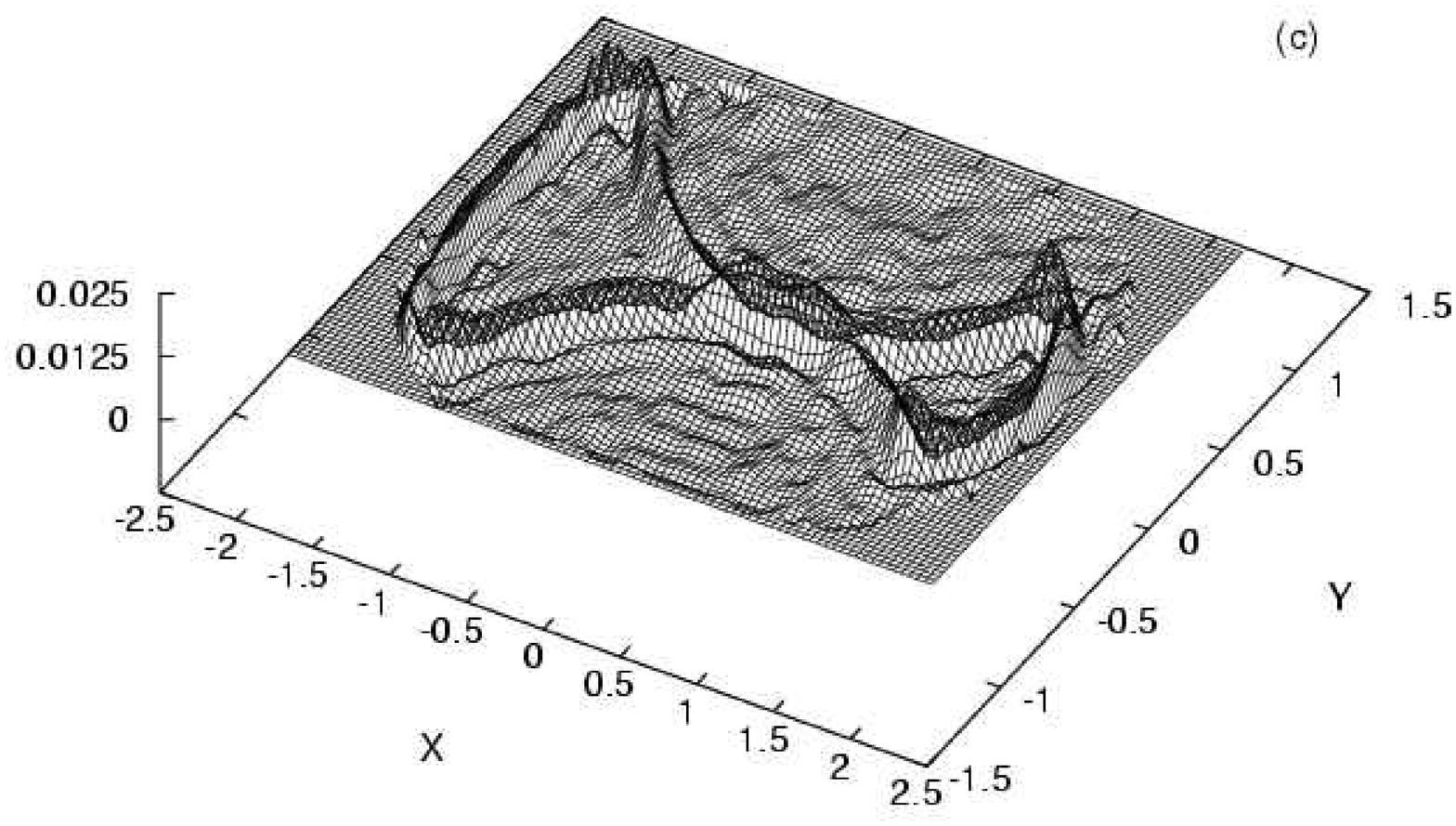}
\includegraphics[width=4.cm,angle=270]{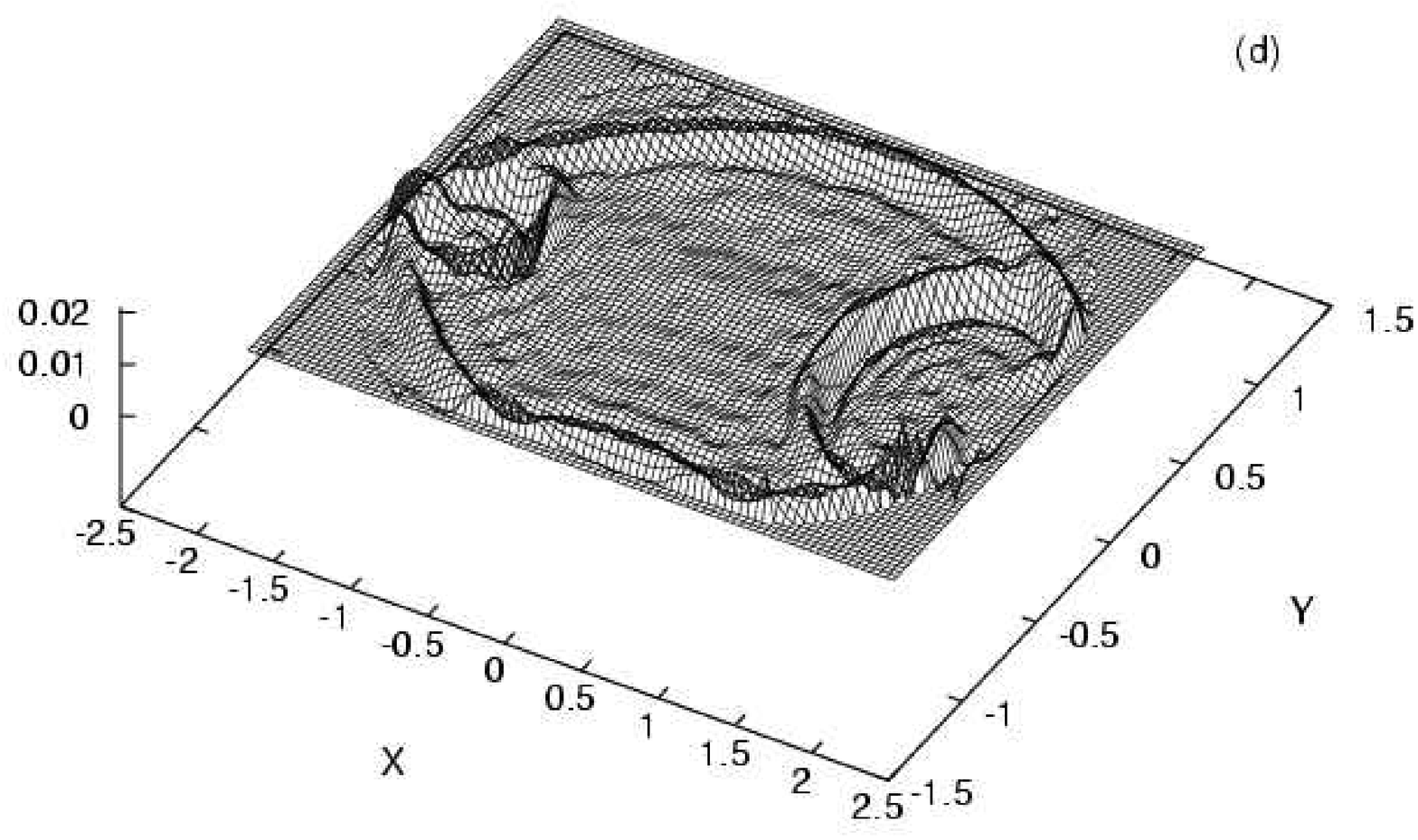}
\includegraphics[width=4.cm,angle=270]{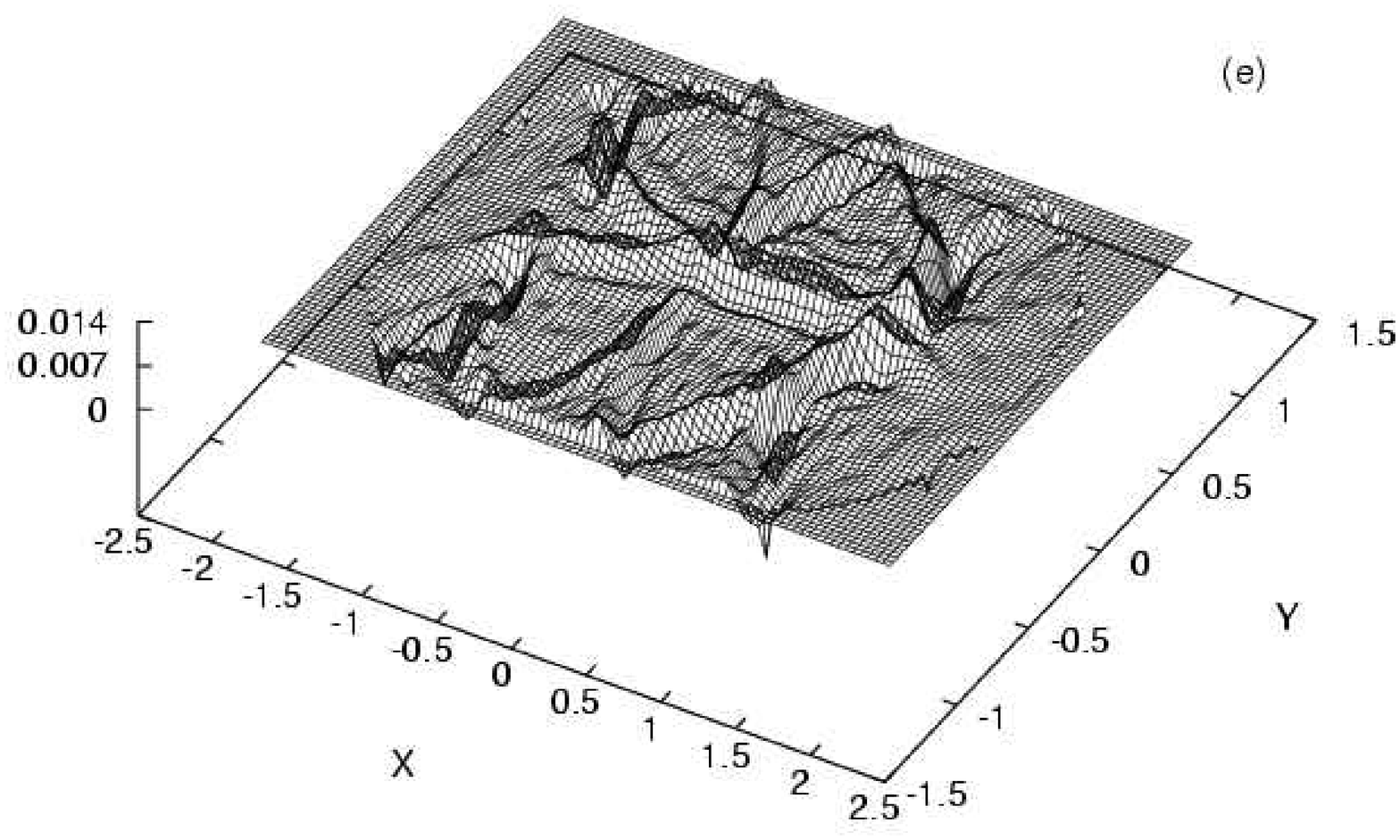}
\includegraphics[width=4.cm,angle=270]{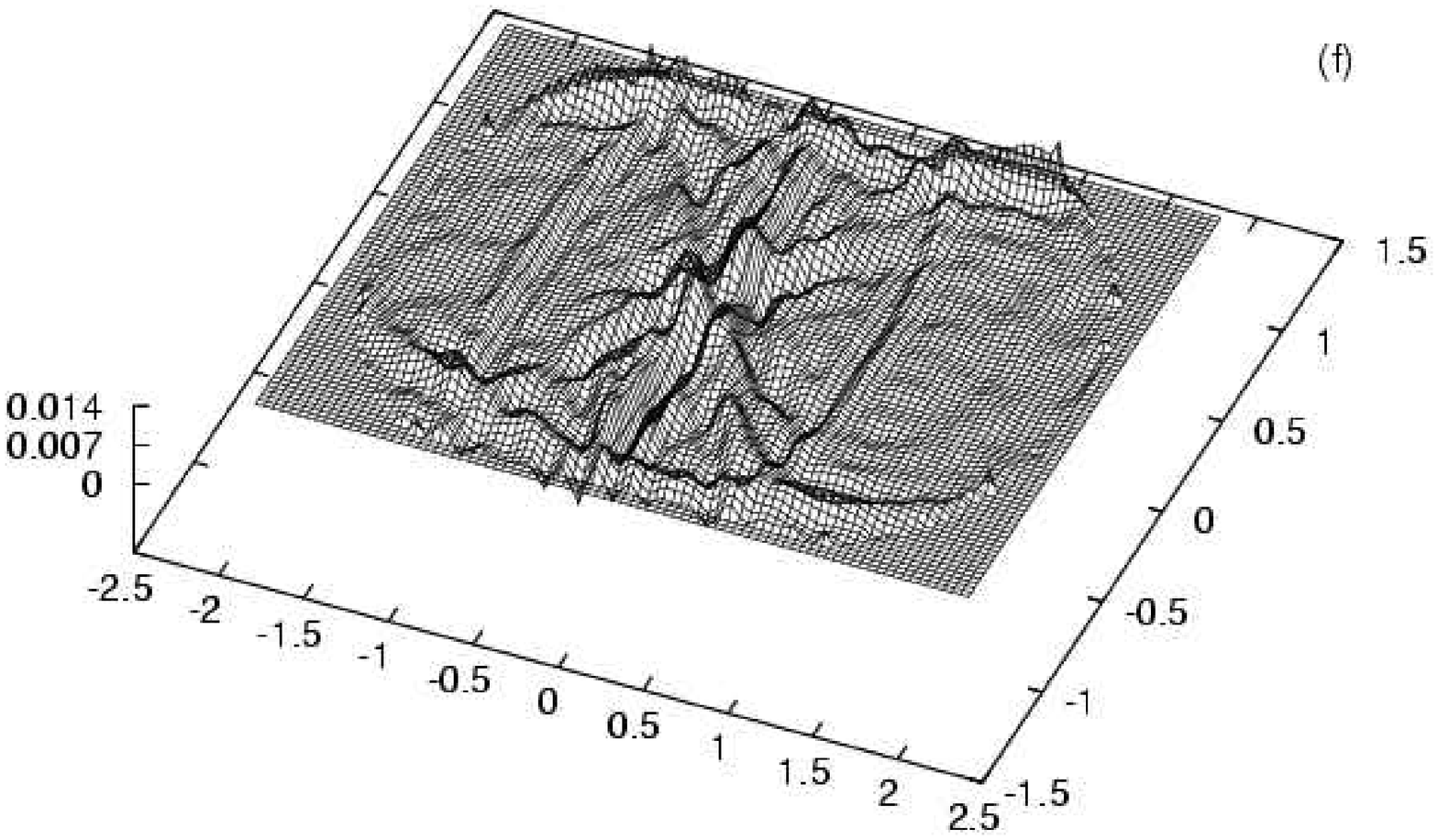}
\end{center}
\caption[ty] {As in Fig.~1 for the stadium billiard at times (a) t=0 (b) t=1 (c) t=2
(d) t=3 (e) t=4 and (f) t=5.
}
\label{fig:stad_quant}
\end{figure*}

\begin{figure*}[tbp]
\begin{center}
\includegraphics[width=4.cm,angle=270]{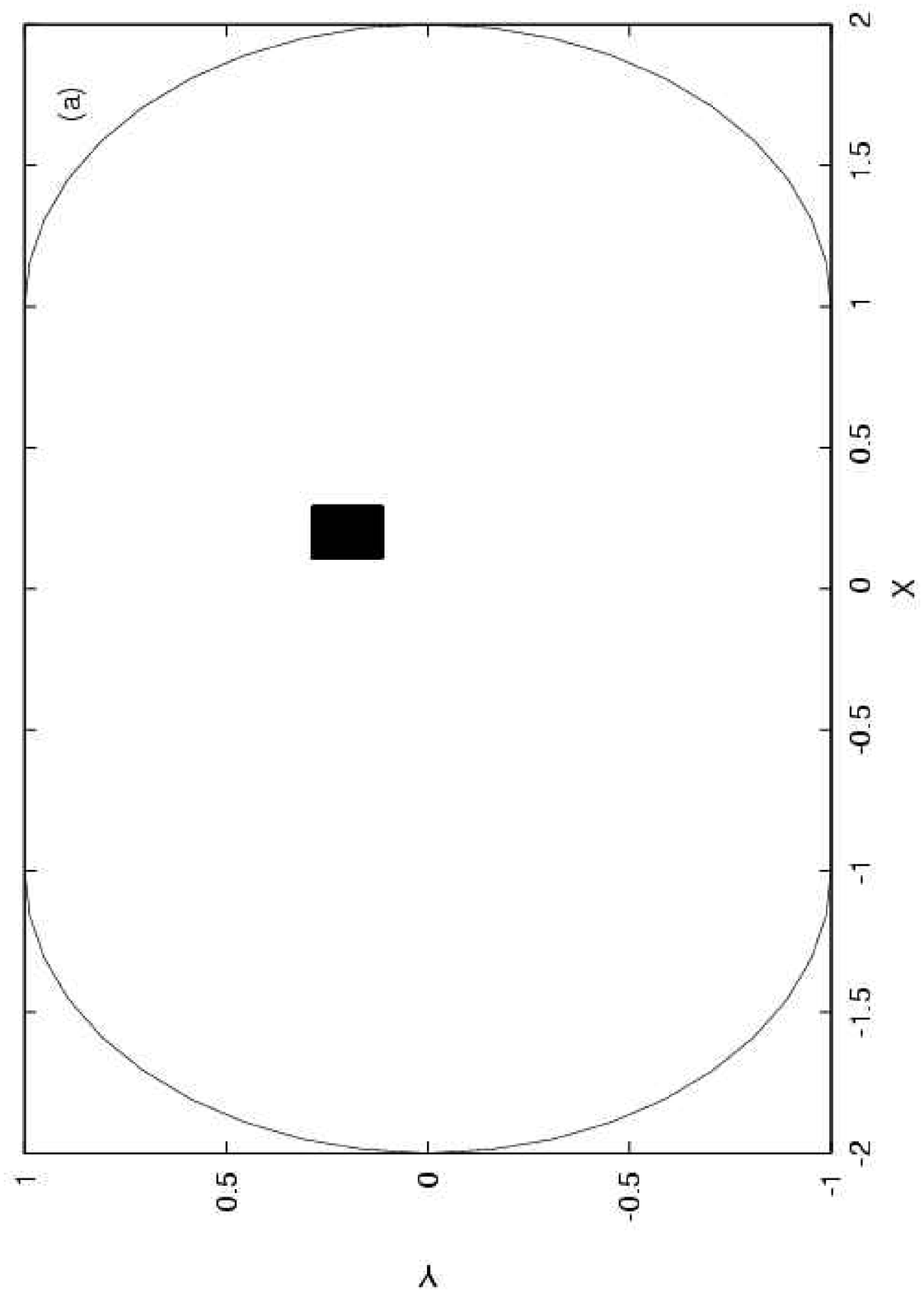}
\includegraphics[width=4.cm,angle=270]{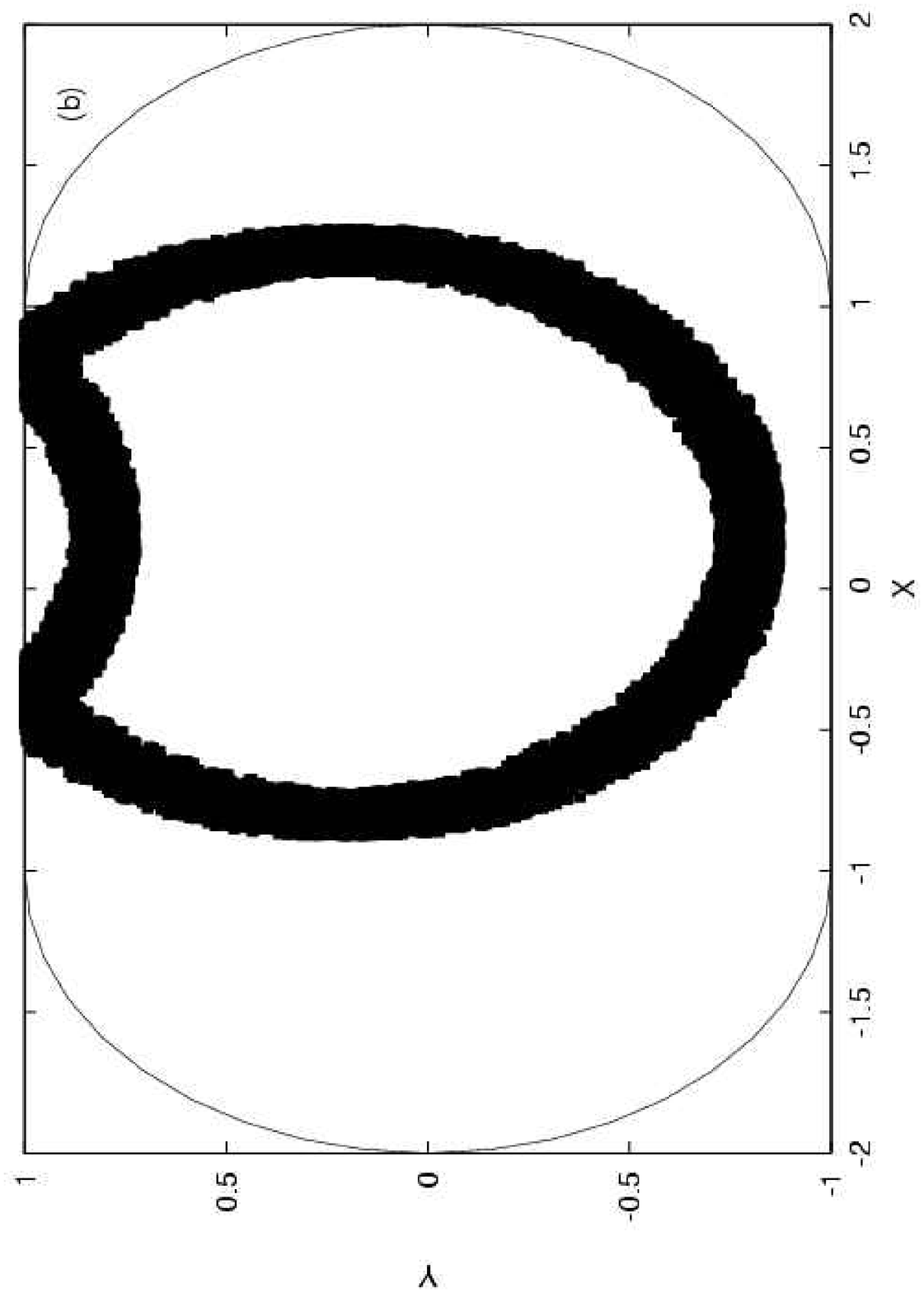}
\includegraphics[width=4.cm,angle=270]{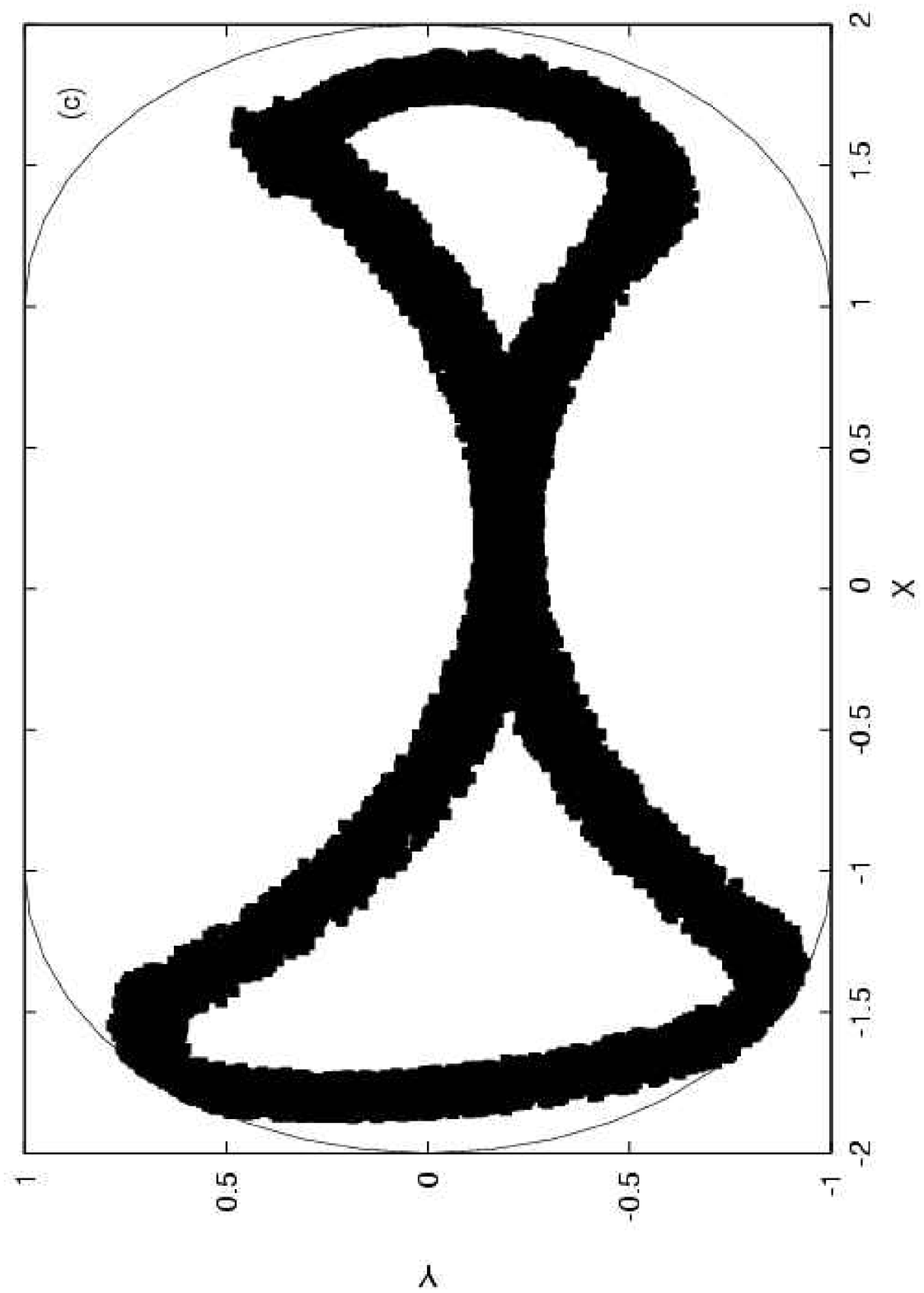}
\includegraphics[width=4.cm,angle=270]{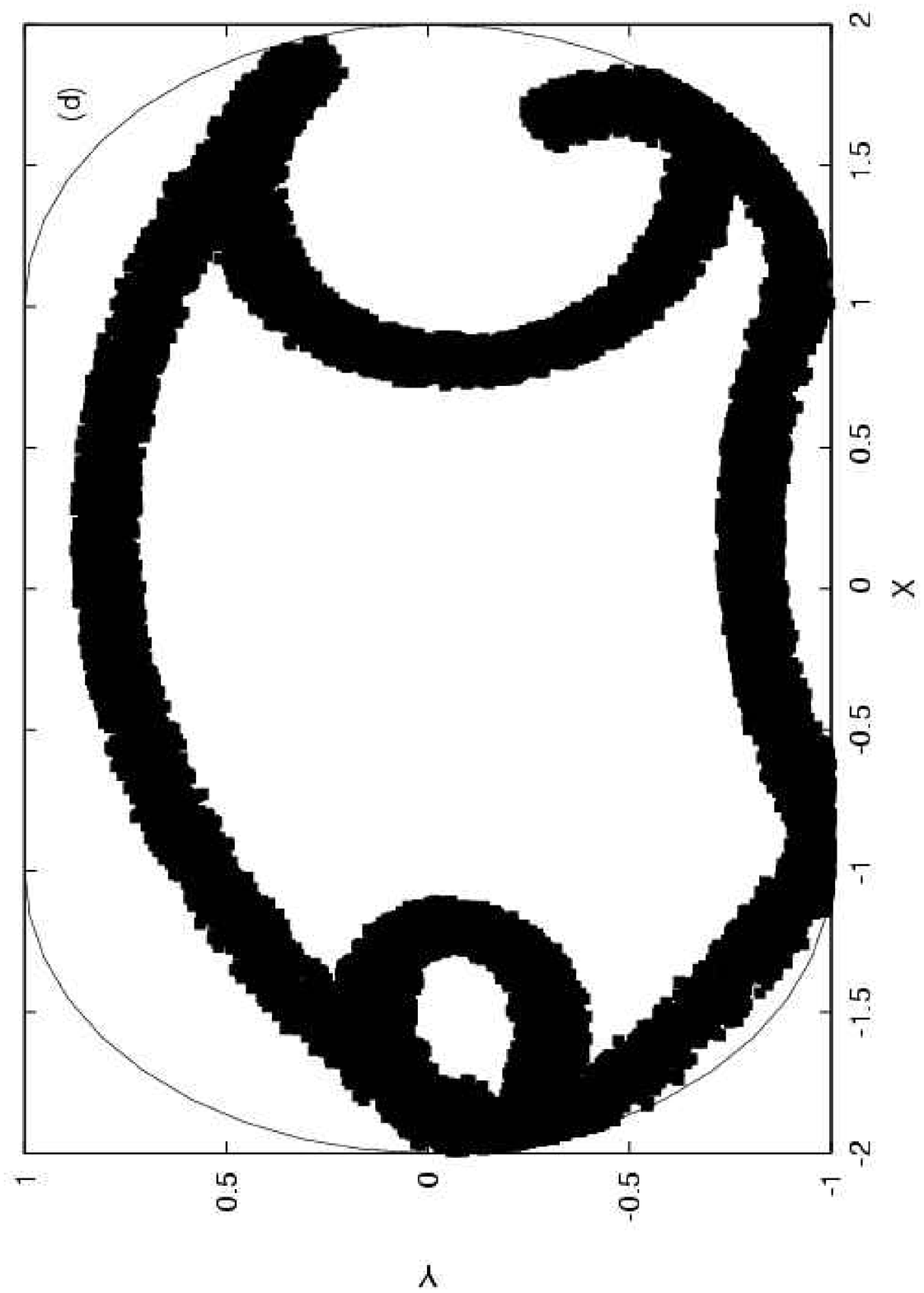}
\includegraphics[width=4.cm,angle=270]{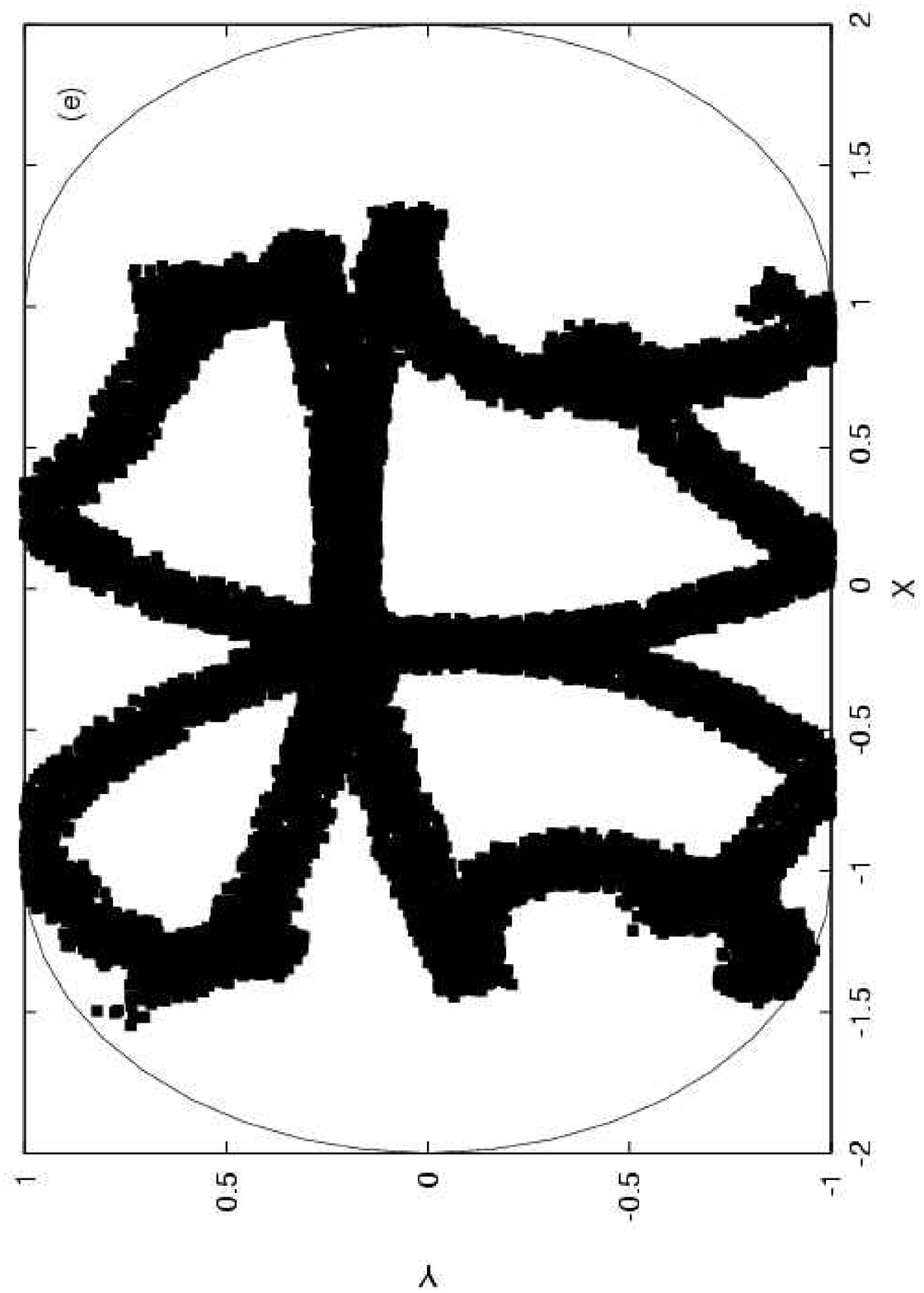}
\includegraphics[width=4.cm,angle=270]{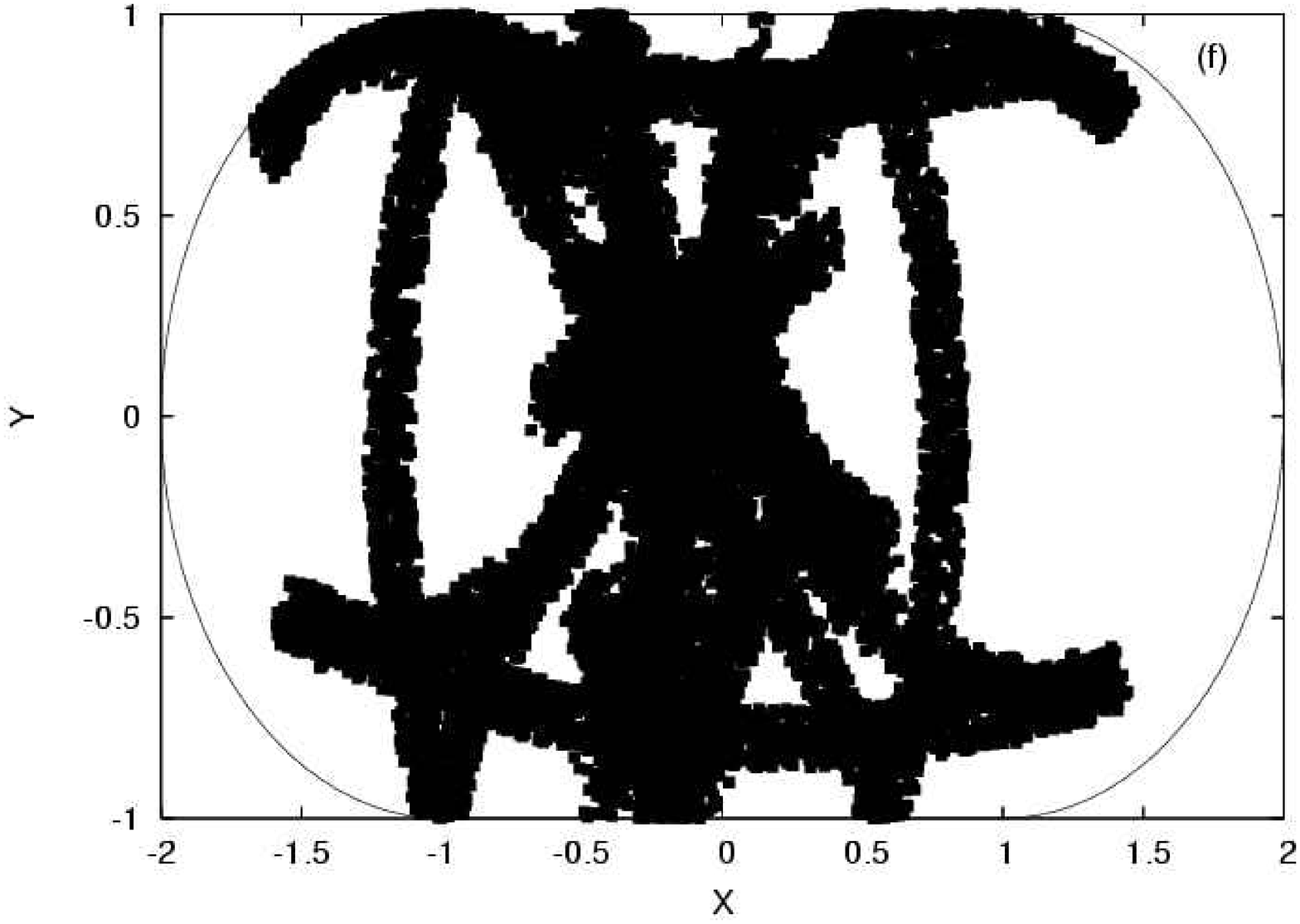}
\end{center}
\caption[ty] {As in Fig.~2 for the stadium billiard at times (a) t=0 (b) t=1 (c) t=2
(d) t=3 (e) t=4 and (f) t=5.
}
\label{fig:stad_classical}
\end{figure*}

We next consider the chaotic Stadium billiard consisting
of two parallel straight segments of length $2$ joined on either end
by a semicircle of unit radius. The initial density is again 
centred at (0.2,0.2) with $\Delta L = 0.15$. In this case, 
the quantum Neumann eigenfunctions
have been computed numerically using the boundary integral method. 
We use the first 1000 eigenfunctions to expand the initial density 
and evolve it using Eq.~(\ref{eq:evolve}). Fig.~\ref{fig:stad_quant} is similar to 
Fig.~\ref{fig:rect_quant}
but for times t=0,1,2,3,4 and 5 while Fig.~\ref{fig:stad_classical} 
(similar to Fig.~\ref{fig:rect_classical}) 
shows the classical evolution of the initial points at these times. 
The similarity between the two evolutions is again obvious despite
the approximate nature of the eigenfunctions and eigenvalues used.

\begin{figure*}[tbp]
\begin{center}
\includegraphics[width=5.cm,angle=270]{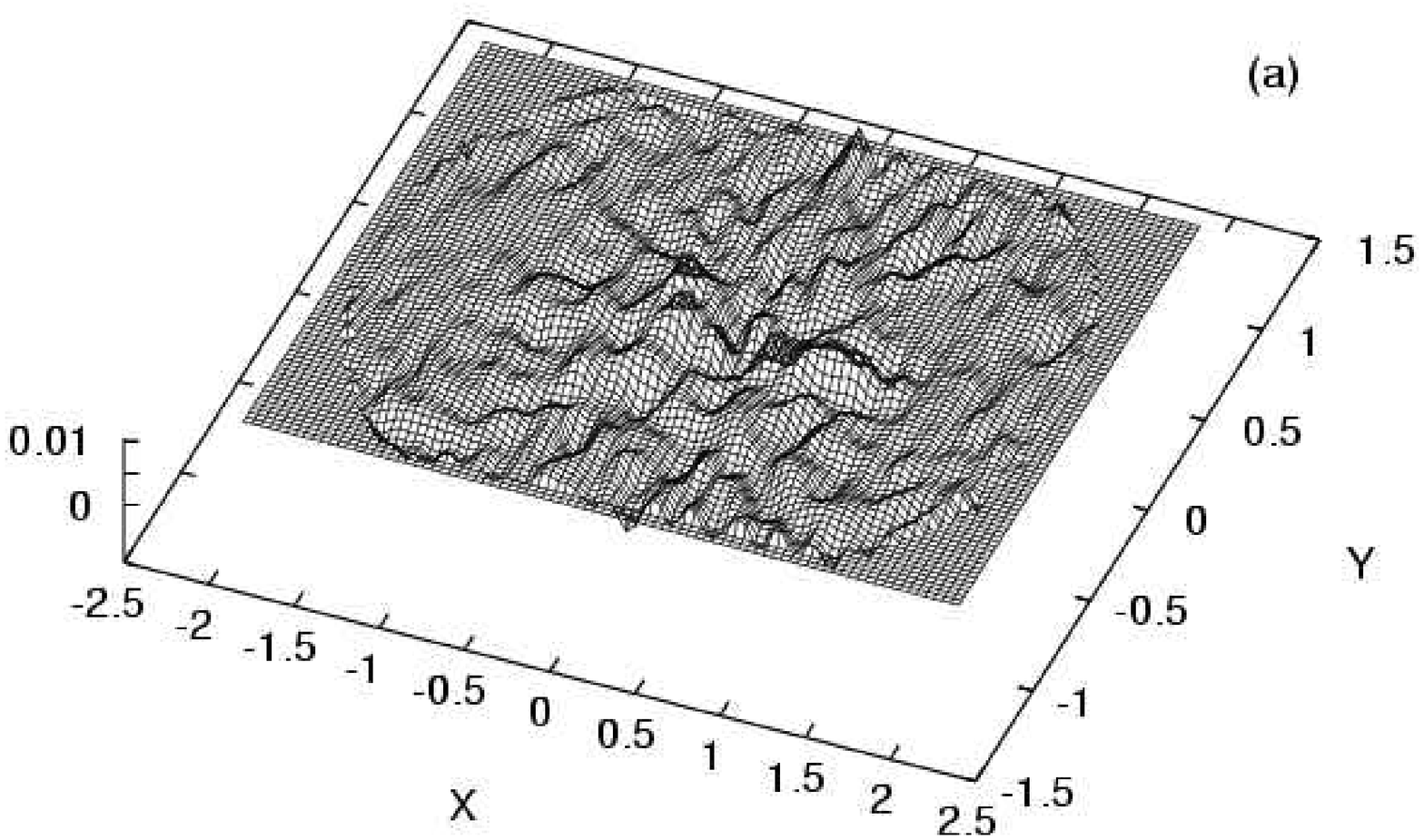}
\includegraphics[width=5.cm,angle=270]{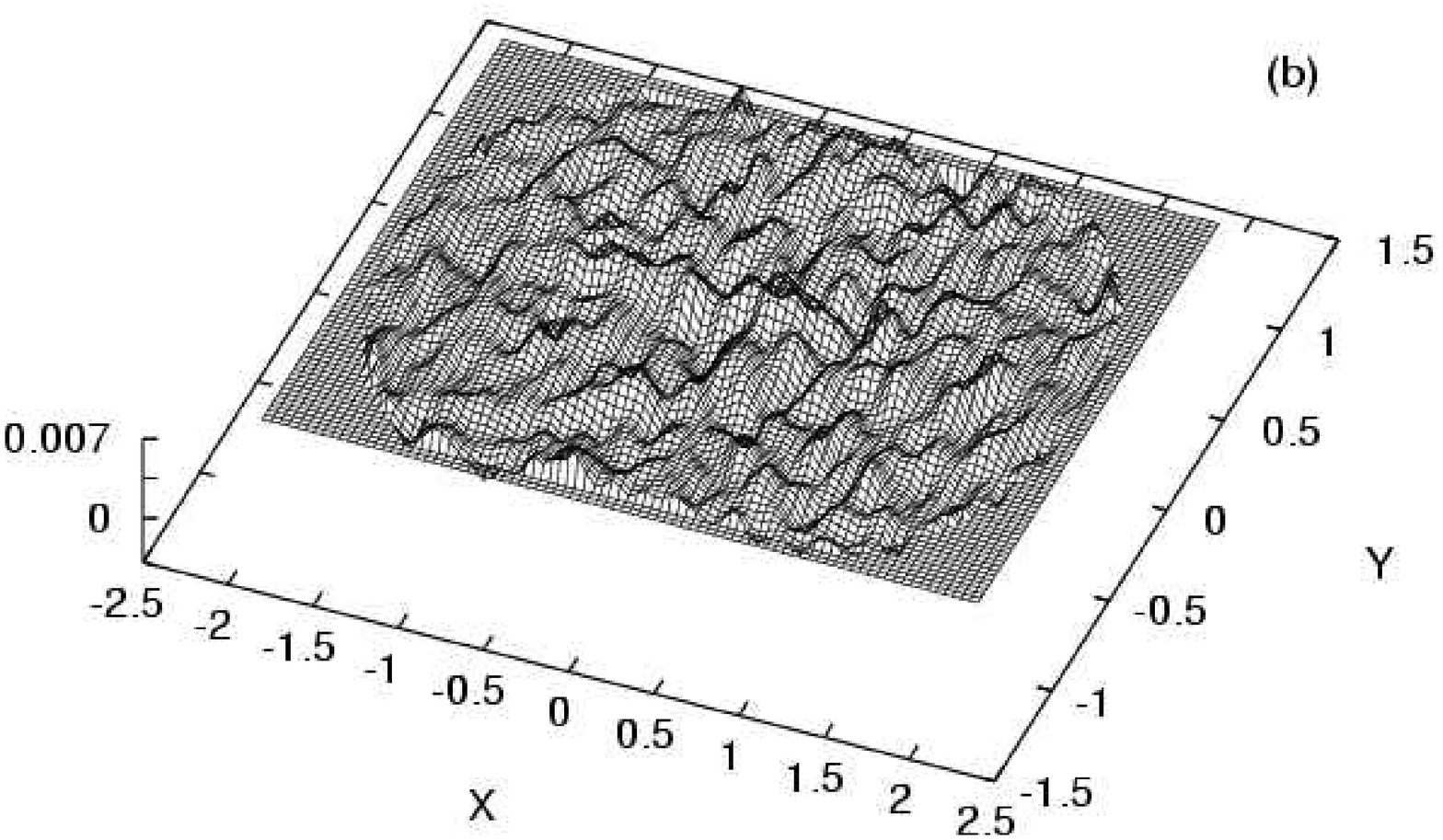}
\includegraphics[width=5.cm,angle=270]{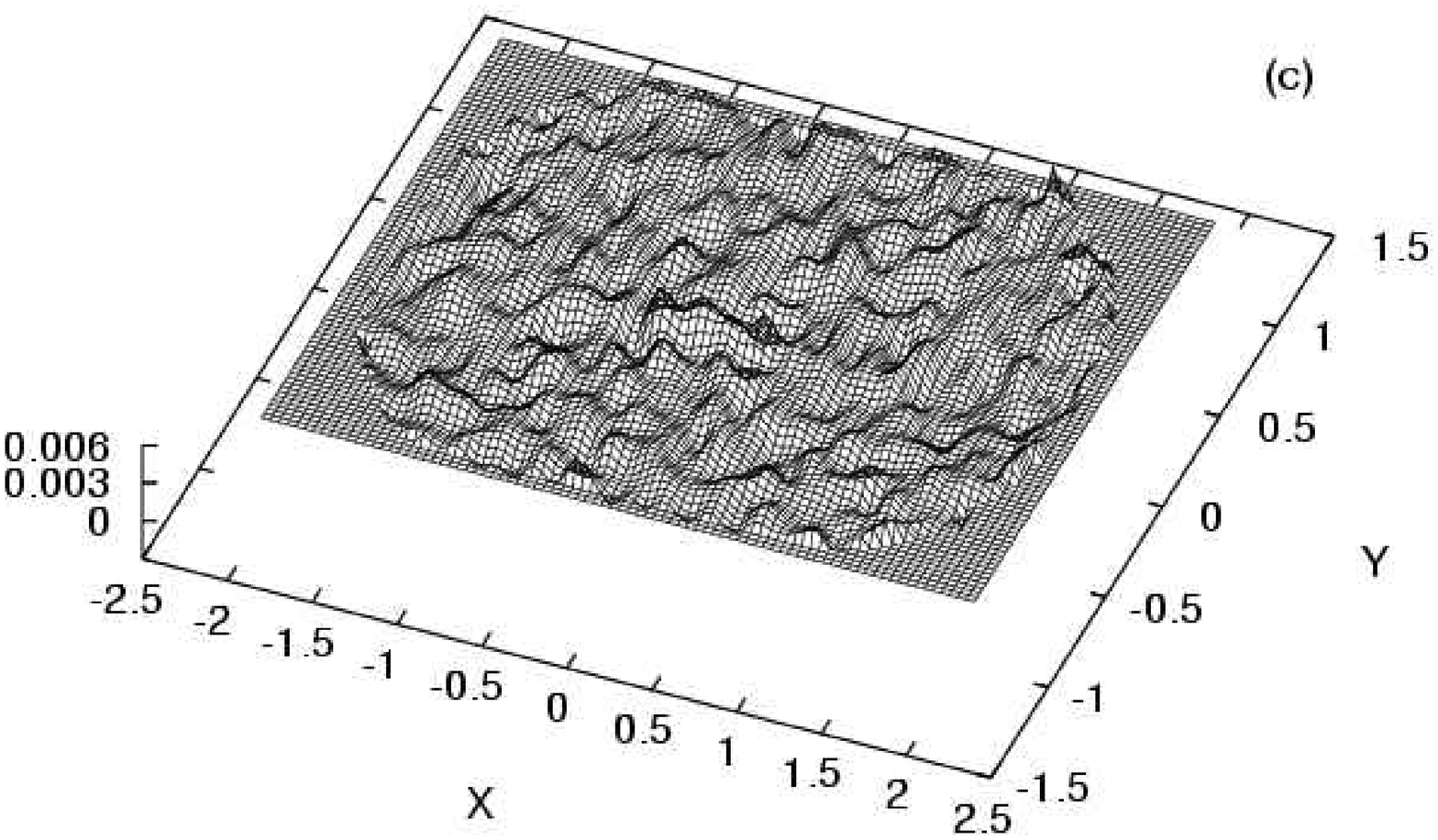}
\end{center}
\caption[ty] {The density evolved using Eq.~\ref{eq:evolve} 
for the stadium billiard at times (a) t=10 (b) t=20 and (c) t=30.
}
\label{fig:stad_quant_long}
\end{figure*}

\begin{figure*}[tbp]
\begin{center}
\includegraphics[width=5.cm,angle=270]{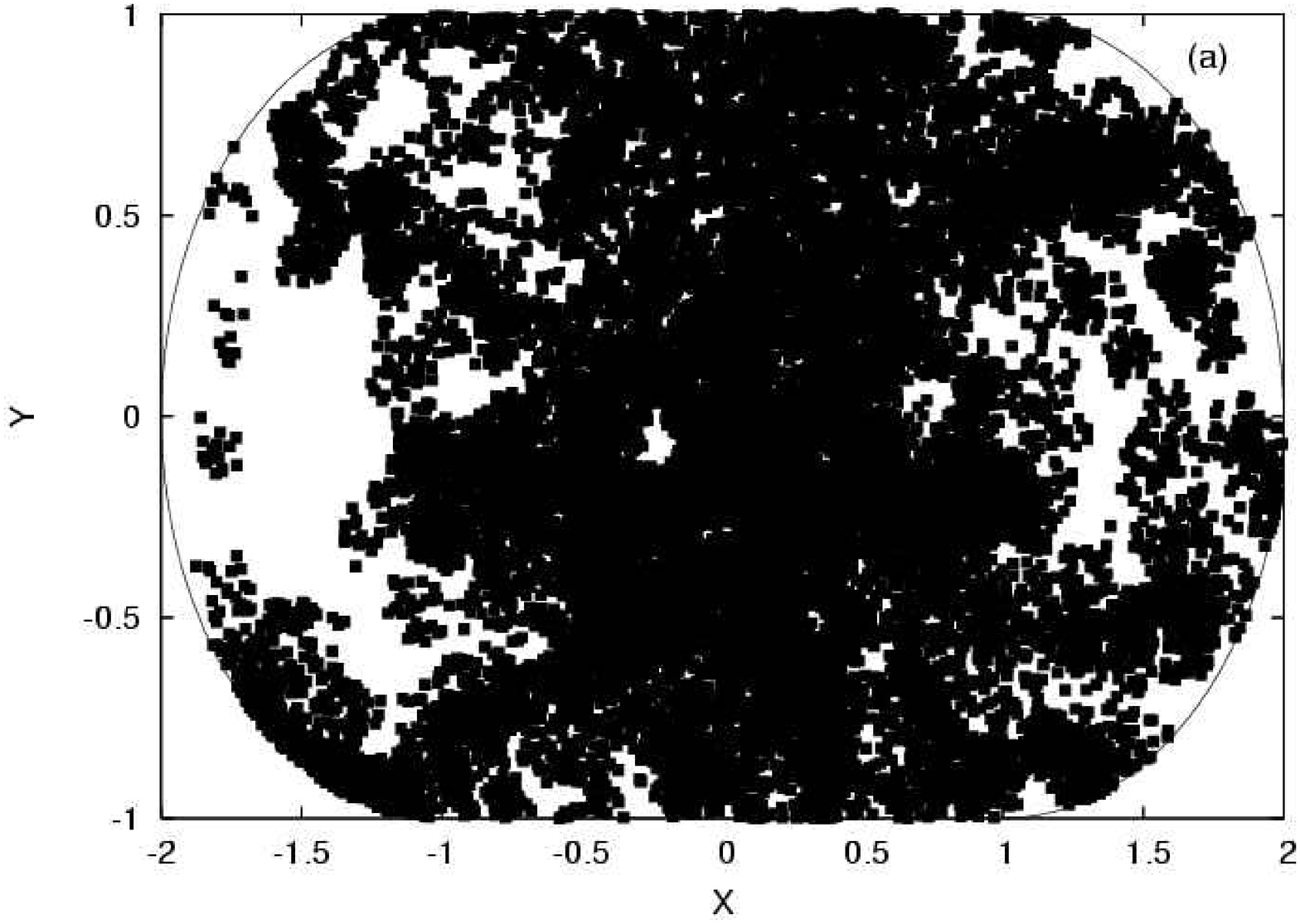}
\includegraphics[width=5.cm,angle=270]{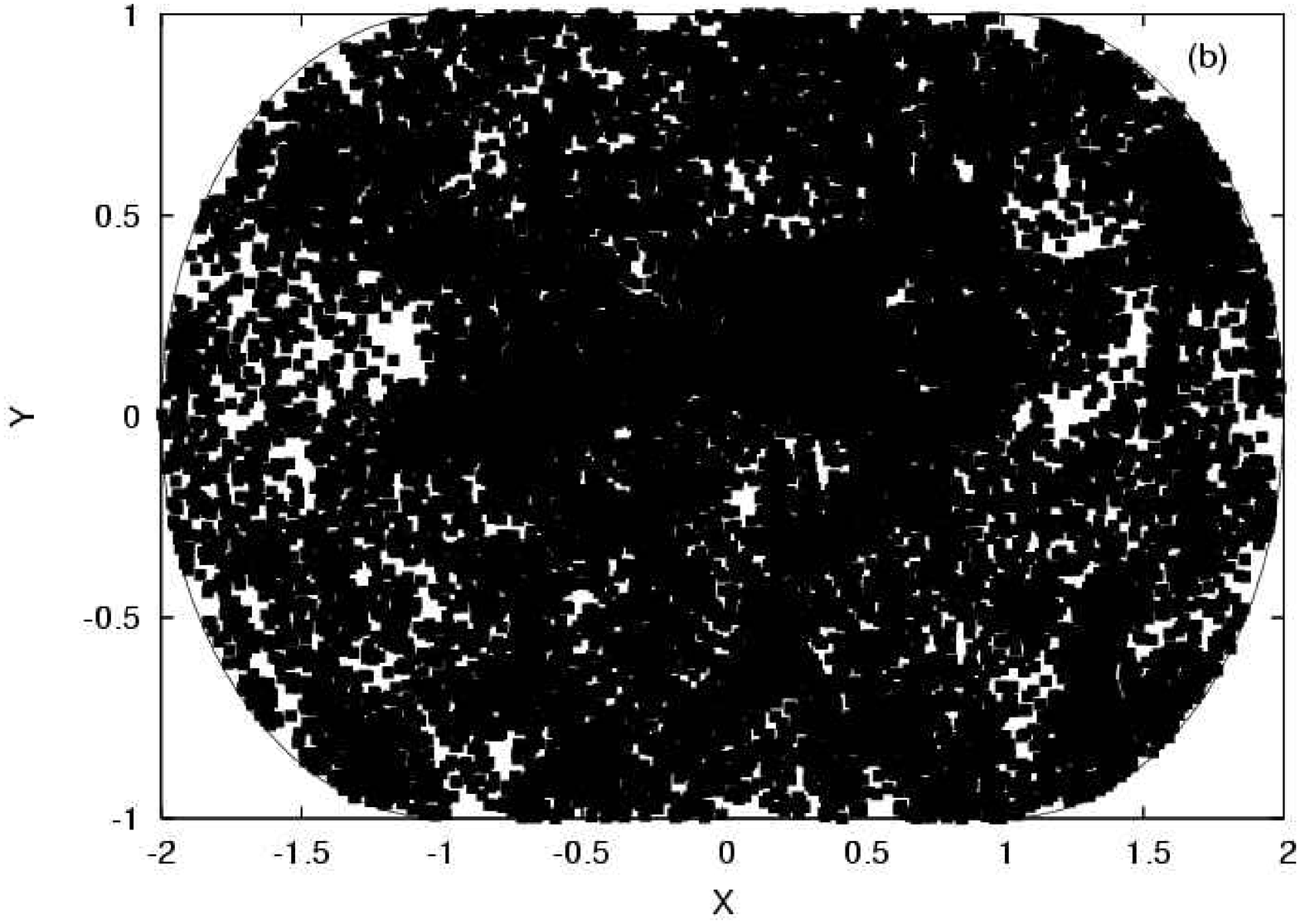}
\includegraphics[width=5.cm,angle=270]{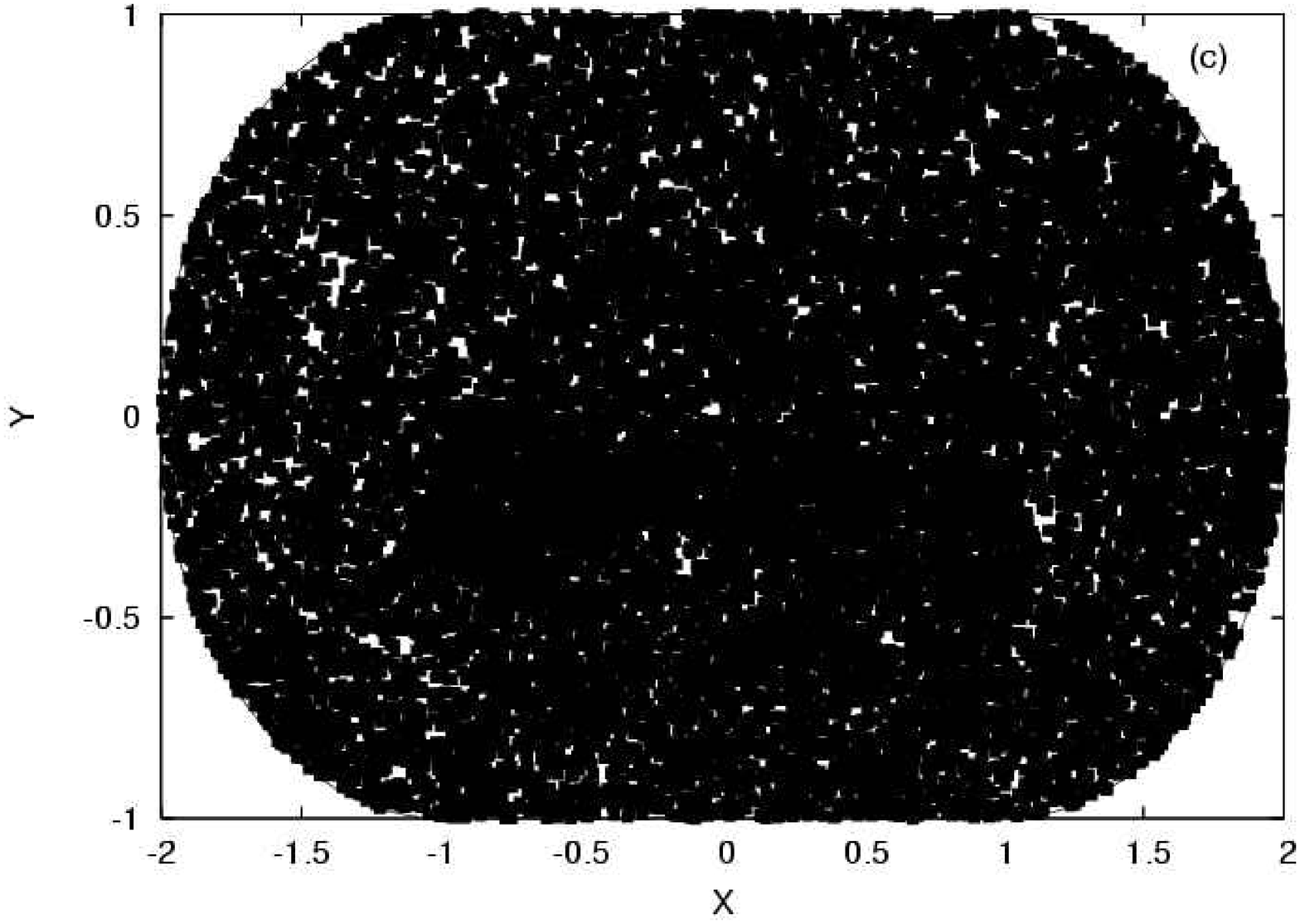}
\end{center}
\caption[ty] {The initial density of Fig.~4a evolved classically using momentum directions
 distributed uniformly  at times (a) t=10 (b) t=20 and (c) t=30.
}
\label{fig:stad_classical_long}
\end{figure*}

At longer times, both evolutions lead towards a uniform distribution in ${\bm q}$ space. 
Fig~\ref{fig:stad_quant_long} shows the density evolved using 
Eq.~\ref{eq:evolve} at times t=10, 20 and 30 corresponding on an average to 4.5, 
9.0 and 13.6 bounces. This is compared to the classical evolution of the initial 
points as these times in Fig.~\ref{fig:stad_classical_long}. Note that
the final invariant density should assume a value $\rho \simeq 0.00315$  for the
initial density considered. Clearly, the density evolution using Eq.~\ref{eq:evolve} 
approaches this value with 
fluctuations around it that  
diminish with time (see Fig.~\ref{fig:stad_quant_long}).  
At $t=30$, the density appears to be nearly uniform 
in both cases (Fig.~(\ref{fig:stad_quant_long}c) and Fig.~(\ref{fig:stad_classical_long}c)). 
The decay to the correct (uniform) invariant density is not surprising as
the Neumann ground state eigenfunction is uniform and the corresponding
eigenvalue $E_0 = 0$ so that $J_0(\sqrt{E_0}vt) = 1$. 
However, the finer structures are harder to discern at longer times. 

\begin{figure*}[tbp]
\begin{center}
\includegraphics[width=14cm,angle=270]{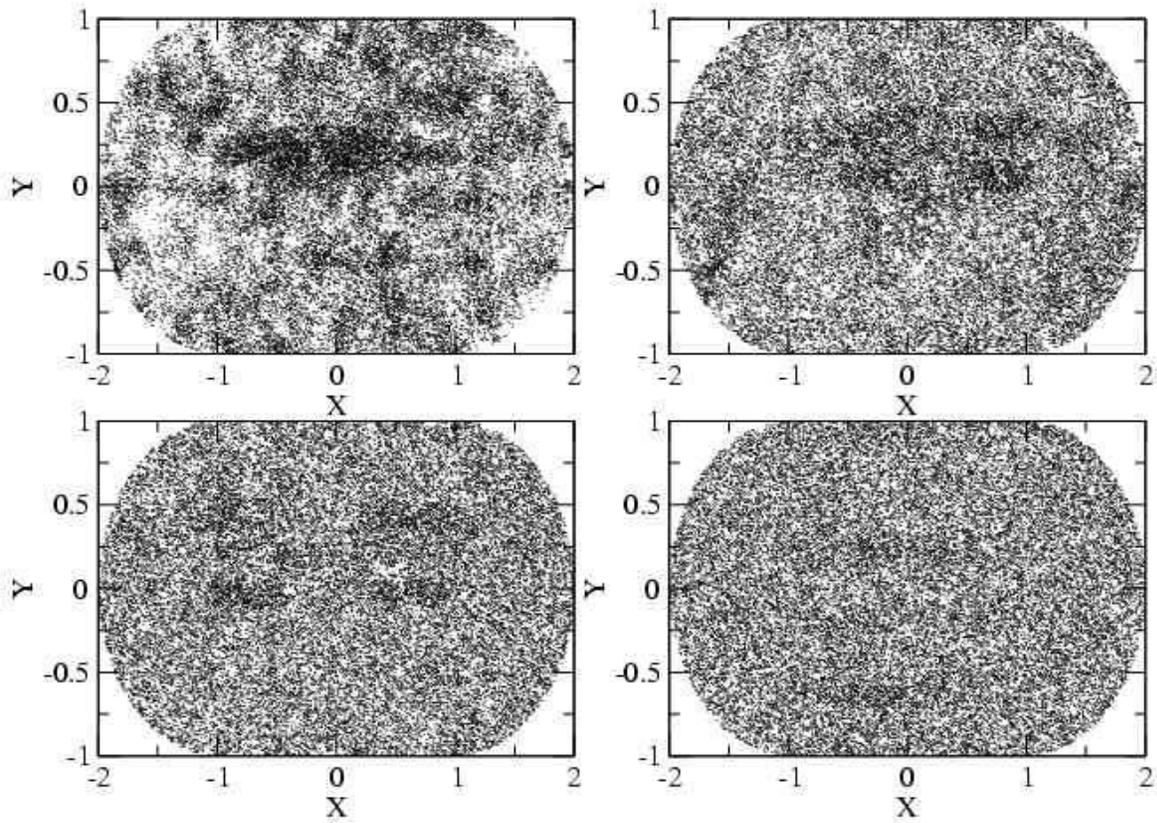}
\end{center}
\caption[ty] {The initial density of Fig.~4a evolved classically using momentum directions
 distributed uniformly  at times  t=20 (top-left),  t=40 (top-right),  t=60 (bottom-left) and (bottom-right) t=110(bottom-right) . 
}
\label{fig:stad_longer_hat}
\end{figure*}

\begin{figure*}[tbp]
\begin{center}
\includegraphics[width=14cm,angle=270]{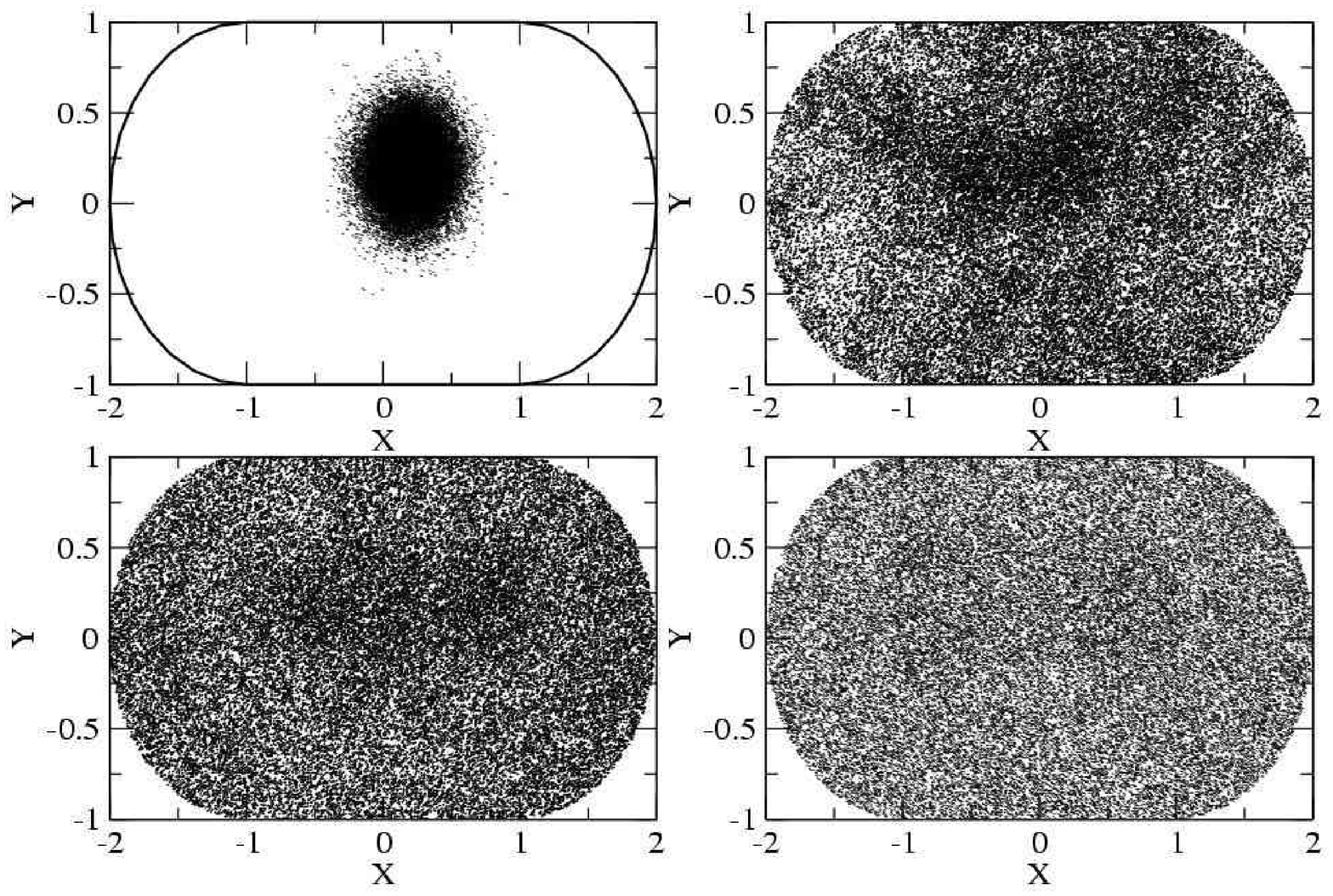}
\end{center}
\caption[ty] {An initial gaussian density (in ${\bm q}$ is evolved classically using momentum directions
 distributed uniformly  at times  t=0 (top-left),  t=20 (top-right),  t=40 (bottom-left) and (bottom-right) t=60(bottom-right) . 
}
\label{fig:stad_longer_gaussian}
\end{figure*}

A closer inspection of the classical evolution however shows that that the
density is nonuniform even at t =  30.  Fig~\ref{fig:stad_longer_hat} shows
the evolution at t = 20, 40, 60 and 110 using a finer representation (dot size) of a
point. In the first two cases, the departure from uniformity seems evident
while at t = 60 (average of 27 bounces), a closer inspection shows the presence of four patches of 
higher density. Finally at t = 110, the density appears to be uniform
corresponding to an average of 50 bounces per trajectory.
The classical evolution of a Gaussian projected density centred at ${\bm q} = (0.2,0.2)$
is shown in Fig.~\ref{fig:stad_longer_gaussian} at t = 0, 20, 40 and 60. Note that at t = 40,
the density is non-uniform while at t = 60 (average of 27 bounces),
the density appears to be uniform. The decay rate thus depends on the initial projected density.

There are two sources of errors in the stadium billiard evolution. 
In the first place, the quantum Neumann eigenfunctions are approximate classical 
eigenfunctions, and, the quantum Neumann eigenvalues $\{E_n\}$ used in $f(\sqrt{E_n}tv)$ 
are close to but not exactly equal to the
classical values. The latter is borne out by the fact that the position of 
peaks in the fourier transform of $K_P$ are
close to but not exactly at the quantum Neumann eigenvalues \cite{arb2}. 
The second source of error arises from the the truncation of the basis.
As associated problem is the fact that the quantum Neumann eigenfunctions
used in the study have been evaluated numerically and hence have errors.
 In the stadium billiard
a thousand eigenstates have been used to expand the initial density 
while in the rectangle (where an analytic expression is used)
five thousand eigenstates have been used to expand and evolve the density. The oscillatory 
background of the initial density in Fig~(\ref{fig:stad_quant}a) is a consequence of truncation
error and the error in numerical evaluation of the exact Neumann eigenfunctions.
This, coupled with the first source of error mentioned above, can even 
lead to negative values of the density at some points.  
 
Despite these problems, evolution using the quantum Neumann eigenfunctions
and the eigenvalues $\{f(\sqrt{E_n}tv)\}$ captures the classical trajectory evolution 
fairly accurately. The finer structures are reproduced well for about 5 bounces 
while at longer times, evolution using the quantum Neumann eigenstates leads to 
the correct invariant density.

\section{Discussion and Conclusions}
\label{sec:discussions}

We have dealt with the evolution within a billiard enclosure of an initial  
classical density on the energy shell that is isotropic in momentum. 
We have constructed an appropriate classical evolution operator ($\Lop_P^t$)
for the momentum projected density and shown that its approximate (sometimes exact)
eigenvalues and eigenfunctions have a one-to-one correspondence with the quantum Neumann
eigenvalues and eigenfunctions. Based on this, we have demonstrated that
for both the
rectangular and stadium billiards, expansion of the 
initial projected-density using the quantum Neumann eigenfunctions as
eigenfunctions of $\Lop_P^t$, 
captures the classical evolution fairly accurately.  
While the finer structures are reproduced 
at shorter times (about 5 bounces for the stadium billiard), 
evolution using the quantum Neumann eigenstates leads to 
the correct invariant density at longer times. Note that the spectrum of the evolution 
operator $\Lop_P^t$ 
does not fix a time scale in which the initial density decays to the  
invariant density.

It is instructive to discuss the difference between evolution due to 
$\Lop_P^t$ and a purely quantum evolution. Specifically, we may ask whether
the hat function that we considered as the initial classical density,
can be evolved quantum mechanically in an identical fashion at least 
for short times.
The initial quantum state $\psi({\bm q},t=0)$ 
in this case is again a hat function (as it is the square root of the initial 
density, $\rho({\bm q})$ ) and its evolution is governed by

\be
\psi({\bm q},t) = \sum_n c_n e^{i E_n t} \psi_n({\bm q},0)
\label{eq:quant_ev}
\ee

\noi
where $c_n = \int_\Delta \psi_n({\bm q},0) d{\bm q}$.  Here $\Delta$ refers
to the region in configuration space where the initial wavefunction
is non-zero. Eq.~\ref{eq:quant_ev} however contains no information 
about the energy or momentum of the underlying classical trajectories.
Thus, evolution due to  Eq.~\ref{eq:quant_ev} cannot approximate classical evolution.
At the semiclassical level, this is due to the fact that trajectories
with all possible energy contribute to the propagator and a correspondence
with classical evolution is possible only when the Wigner transform of
the initial wavepacket is localized in phase space. 
In contrast, evolution due to $\Lop_P^t$ has information about the
magnitude of the momentum ($v$ in the eigenvalue) and the isotropy 
in momentum is built into the kernel $K_P$ and reflects in the form of the eigenvalue 
(the Bessel function $J_0$).

Another important distinction concerns the decay rates 
of initial phase space densities that are directionally (or $\varphi$) dependent and 
independent (isotropic or uniform in $\varphi$).
To explore this further, we shall use the operator ${\Pop}$ to denote 
momentum projection. For the anisotropic or directionally dependent initial
density $\rho(\bm{q}, \bm{p}) = \rho(\bm{q},p,\varphi)$, 

\bea
(\Pop \circ \Lop^t) \rho({\bm q},{\bm p}) 
& = & \Pop \sum_n C_n \Lop^t \phi_n({\bm q},{\bm p}) \nonumber \\ 
& = & \sum_n C_n e^{\alpha_n t} \Pop \phi_n({\bm q},{\bm p}) \nonumber \\
& = & \sum_n C_n e^{\alpha_n t} \tilde{\phi}_n({\bm q}).
\eea 

\noi
Here $\tilde{\phi}_n({\bm q}) = \Pop \phi_n({\bm q},{\bm p})$, 
$\phi_n({\bm q},{\bm p})$ are the eigenfunctions of $\Lop_P^t$ and
the coefficients $C_n$ can be determined using the eigenfunctions
of ${\Lop}^{\dagger}$,  the adjoint of ${\Lop}$. 
If the leading $\alpha_n$ 
has a negative real part as in case of hyperbolic systems, the
decay to the invariant density is exponential.

In contrast, an isotropic initial density  such as 
$\rho(\bm{q}, \bm{p}) = \rho(\bm{q},p) = g(\bm{q})\delta(E_0 - p^2)$
cannot be expanded in eigenfunctions of $\Lop^t$. To understand 
this, note that the only eigenfunction of $\Lop^t$ which is uniform in $\varphi$ (i.e. isotropic) 
is the invariant density which is uniform in both ${\bm q}$ and $\varphi$. All other
densities that are initially uniform in $\varphi$, lose their uniformity (in $\varphi$)
as they evolve with $\Lop^t$. In other words, $\Lop^t$ does not possess any eigenfunction
that is uniform in $\varphi$ but {\em not uniform} in $\bm{q}$. 
Thus $g(\bm{q})\delta(E_0 - p^2)$ cannot be expanded in an eigenbasis of $\Lop^t$. 
The decay to the invariant density  must therefore be
dictated by the eigenvalues of $\Lop_P^t$ since

\be
(\Pop \Lop^t)~g({\bm q}) \delta(E_0 - p^2) =  \Lop_P^t~\rho_P(\bm{q})
\ee

\noi
where $\rho_P({\bm q}) = \int d\bm{p}~ \rho(\bm{q},p) = \pi g(\bm{q})$.
Thus the decay rates of initial densities that are isotropic must differ
from the decay rates of anisotropic densities.

\vskip 0.1 in

A few conclusions 
based on the results of the preceding sections are listed below:

\begin{itemize}

\item{} The evolution of a momentum projected density using the (approximate)
eigenstates of the momentum projected evolution operator $\Lop_P^t$ 
captures the classical evolution fairly accurately. The approximate 
eigenfunctions are the quantum Neumann eigenfunctions while the 
eigenvalues are related to the quantum Neumann eigenvalues.  

\item{} The evolution operator $\Lop_P^t$ is non-multiplicative i.e.
$\Lop_P^{t_2+ t_1} \neq \Lop_P^{t2} \circ \Lop_P^{t_1}$. Thus, the
eigenvalues do not have the form $e^{\lambda_i t}$.

\item{} $\Lop_P^t$ preserves positivity of the
initial density. However numerical errors and the approximate nature
of the eigenvalues and eigenfunctions used in the evolution (Eq.~\ref{eq:evolve})
can lead to negative values at some points. 

\item{}
As the small $t$ behaviour of the density is reproduced by Eq.~(\ref{eq:evolve}),
the correct form of the approximate eigenvalues of $\Lop_P^t$ is
$\{J_0(\sqrt{E_n}vt)\}$ where $E_n$ are the quantum Neumann energy eigenvalues. 
For integrable billiards such as
the rectangle, these are the exact eigenvalues. 

\item{}
For non-polygonal billiards such as the stadium, proof of the existence 
of a correspondence was indirect, based on the limit of a large number of sides (of the
corresponding polygonalized billiard), and, 
numerical evidence based on a few eigenstates \cite{prl2004,pramana2005}. 
The results presented here show that this correspondence must hold for all states
as the evolution of the density using the quantum Neumann eigenstates faithfully follows 
the trajectory picture. This puts the correspondence on a firmer footing.

\item{}
The relaxation of an arbitrary projected-density to the uniform, steady-state density
can be predicted reasonably using the approximate eigenvalues and eigenfunctions
of the projected Perron-Frobenius operator, $\Lop_P^t$.

\end{itemize}

\appendix
\section{}
\label{app}

We shall show here that in polar coordinates ($p,\varphi$),

\be  
\int{\bm \delta}({\bm p} - {\bm p'}^t) h({\bm q'},{\bm p'})  d{\bm p'} 
= \int d\varphi'~\delta(\varphi - \varphi'^t) h({\bm q'},p,\varphi').
\ee

\noi
where $\varphi'^t = \varphi^t({\bm q'},p,\varphi')$ is the polar angle
of the momentum vector ${\bm p'}$
at time $t$ for an initial phase space point (${\bm q'},p,\varphi'$).
Suppressing ${\bm q'}$, the integral can be expressed as

\be
\int \delta(p\cos\varphi - p'\cos\varphi'^t) \delta(p\sin\varphi - p'\sin\varphi'^t)h({\bm p'}) 
d{\bm p'} 
\label{eq:app_polar}
\ee

\noi 
where it is assumed that for billiards $p'^t = p'$. Obviously the only value of ${\bm p'}$
that contributes to  the integral is $p' = p$ and $\varphi'^t = \varphi$. In order to 
carry out the $p'$ integration, we may rewrite (\ref{eq:app_polar}) as

\be
\int {\delta(p{\cos\varphi\over\cos\varphi'^t} - p') \over \cos\varphi'^t} 
\delta(p\sin\varphi - p'\sin\varphi'^t) h({\bm p'}) 
d{\bm p'} 
\ee

\noi
On performing the $p'$ integration, we have

\bea
& \; & \int d\varphi'~\delta(p\sin\varphi - p\cos\varphi \tan\varphi'^t) 
{p \cos\varphi\over \cos^2\varphi'^t} h(p,\varphi')\nonumber \\
& = & \int d\varphi'~{\delta(\tan\varphi - \tan\varphi'^t) \over \cos^2\varphi'^t}h(p,\varphi')
 \nonumber \\
& = & \int d\varphi'~\delta(\varphi - \varphi'^t) h(p,\varphi').
\eea

\section{Non-multiplicative property of $\Lop_P^t$}
\label{app2}

The evolution of momentum projected isotropic initial densities on the energy shell 
is governed by $\Lop_P^t$:

\be
\Lop_P^t \circ  \rho_P(\bm{q}) = \int d\bm{q}'~K_P(\bm{q},\bm{q}',t)~\rho_P(\bm{q}').
\ee

\noi
The non-multiplicative property of $\Lop_P^t$ implies
 
\be
\int~d{\bm q''}~K_P({\bm q},{\bm q''},t_2)~K_P({\bm q''},{\bm q'},t_1) \neq
K_P({\bm q},{\bm q'},t_1 + t_2)
\ee

\noi
This is easiest to visualize when the points ${\bm q'}$, ${\bm q}$ and the times $t_1$ and $t_2$  
are such that such that there is no encounter (reflection) 
with the boundary of the billiard. As the momentum vector
does not change direction, $\varphi^* = \varphi$ in Eq.~\ref{eq:kernel}. 
Using Eq.~\ref{eq:kernel} for the projected
kernel $K_P$, 

\vskip -0.15 in
\bea
& ~ & \int d{\bm q''}~K_P({\bm q},{\bm q''},t_2)~K_P({\bm q''},{\bm q'},t_1) \nonumber \\
& = & {1\over (2\pi)^2} \int d\varphi d\varphi' d{\bm q''}~ \delta({\bm q} - 
{\bm f_q}^{t_2}({\bm q''},p_0,\varphi')) 
\nonumber \\
& ~ & ~~~~~~~~~ \times \delta({\bm q''} - {\bm f_q}^{t_1}({\bm q'},p_0,\varphi)) \\
& = &  {1\over (2\pi)^2} \int d\varphi d\varphi'~
\delta({\bm q} - {\bm f_q}^{t_2}({\bm f_q}^{t_1}({\bm q'},\varphi),\varphi')) \label{eq:prod} \\
& \neq & {1 \over 2\pi} \int d\varphi~ \delta({\bm q} - {\bm f_q^{t_1+t_2}}({\bm q'},p_0,\varphi))
\eea

\noi
where ${\bm f_q}$ (also denoted by ${\bm q}^t$ in the text) is the ${\bm q}$-component of the flow $f^t$.
 
\begin{figure*}[tbp]
\begin{center}
\includegraphics[width=6.5cm,angle=0]{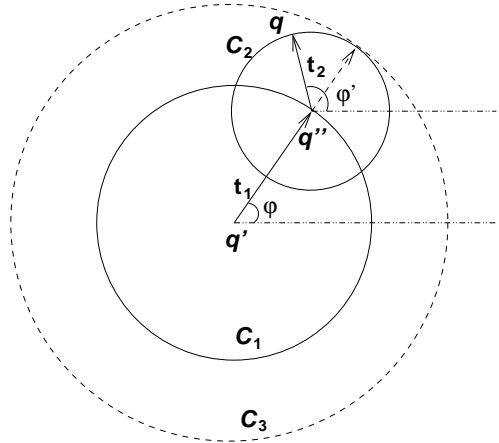}
\caption{Comparison of two time evolutions in a billiard enclosure (boundary not shown).}
\label{fig:kernel}
\end{center}
\end{figure*}

The last step can be best understood in terms of Fig.~\ref{fig:kernel} which 
illustrates the difference  in the two time evolutions.
The first is a two step evolution from ${\bm q'}$ to ${\bm q''}$ in time $t_1$ followed 
by an evolution from ${\bm q''}$  to ${\bm q}$ in time $t_2$. These are indicated 
by solid lines terminated by arrows. The kernel determines the angles $\varphi$ and
$\varphi'$ which connect ${\bm q'}$ with ${\bm q}$ via the intermediate point ${\bm q''}$.
Depending on the angle $\varphi$, the point ${\bm q''}$ can lie anywhere on the inner 
circle ($C_1$). Any point ${\bm q}$ on the circle $C_2$ (with centre on $C_1$) has a contribution
at time $t_1 + t_2$ from the initial density at ${\bm q'}$.
The second evolution is for a time $t_1 + t_2$ starting from ${\bm q'}$. The kernel
for this evolution contributes when ${\bm q}$ lies on the outer circle ($C_3$).

Thus, $\Lop_P^{t_1} \circ \Lop_P^{t_2} \neq \Lop_P^{t_1 + t_2}$. A consequence
of the non-multiplicative nature of $\Lop_P^t$ is that its 
eigenvalues cannot be of the form $e^{\lambda_i t}$ i.e. the eigenvalues of $\Lop^t$
are distinct from the eigenvalues of $\Lop_P^t$ when the initial phase space density is isotropic
and constrained to the constant energy surface.

\newcommand{\PR}[1]{{Phys.\ Rep.}\/ {\bf #1}}
\newcommand{\PRL}[1]{{Phys.\ Rev.\ Lett.}\/ {\bf #1}}
\newcommand{\PRA}[1]{{Phys.\ Rev.\ A}\/ {\bf #1}}
\newcommand{\PRB}[1]{{Phys.\ Rev.\ B}\/ {\bf #1}}
\newcommand{\PRD}[1]{{Phys.\ Rev.\ D}\/ {\bf #1}}
\newcommand{\PRE}[1]{{Phys.\ Rev.\ E}\/ {\bf #1}}
\newcommand{\JPA}[1]{{J.\ Phys.\ A}\/ {\bf #1}}
\newcommand{\JPB}[1]{{J.\ Phys.\ B}\/ {\bf #1}}
\newcommand{\JCP}[1]{{J.\ Chem.\ Phys.}\/ {\bf #1}}
\newcommand{\JPC}[1]{{J.\ Phys.\ Chem.}\/ {\bf #1}}
\newcommand{\JMP}[1]{{J.\ Math.\ Phys.}\/ {\bf #1}}
\newcommand{\JSP}[1]{{J.\ Stat.\ Phys.}\/ {\bf #1}}
\newcommand{\AP}[1]{{Ann.\ Phys.}\/ {\bf #1}}
\newcommand{\PLB}[1]{{Phys.\ Lett.\ B}\/ {\bf #1}}
\newcommand{\PLA}[1]{{Phys.\ Lett.\ A}\/ {\bf #1}}
\newcommand{\PD}[1]{{Physica D}\/ {\bf #1}}
\newcommand{\NPB}[1]{{Nucl.\ Phys.\ B}\/ {\bf #1}}
\newcommand{\INCB}[1]{{Il Nuov.\ Cim.\ B}\/ {\bf #1}}
\newcommand{\JETP}[1]{{Sov.\ Phys.\ JETP}\/ {\bf #1}}
\newcommand{\JETPL}[1]{{JETP Lett.\ }\/ {\bf #1}}
\newcommand{\RMS}[1]{{Russ.\ Math.\ Surv.}\/ {\bf #1}}
\newcommand{\USSR}[1]{{Math.\ USSR.\ Sb.}\/ {\bf #1}}
\newcommand{\PST}[1]{{Phys.\ Scripta T}\/ {\bf #1}}
\newcommand{\CM}[1]{{Cont.\ Math.}\/ {\bf #1}}
\newcommand{\JMPA}[1]{{J.\ Math.\ Pure Appl.}\/ {\bf #1}}
\newcommand{\CMP}[1]{{Comm.\ Math.\ Phys.}\/ {\bf #1}}
\newcommand{\PRS}[1]{{Proc.\ R.\ Soc. Lond.\ A}\/ {\bf #1}}


\clearpage

\begin{thebibliography}{0}
\bibitem{not_the_same} These are not the same as the eigenvalues and eigenfunctions of $\Lop^t$.
See section \ref{sec:projected} and appendix \ref{app2} for non-multiplicative 
property of $\Lop_P^t$, section \ref{sec:estates} for an illustration using 
integrable billiards and the discussion in the concluding section.
\bibitem{planar} In polygonalized billiards, $\varphi'^t({\bm q}',\varphi',p) = \varphi'$
for a trajectory in the unfolded space which is produced by 
successive reflections of the billiard domain about the edge that the trajectory
strikes. Thus in the unfolded space, a trajectory is a straight line.
\bibitem{fnote_momentum} For a better comparision, for each initial point in ${\bm q}$, 
an isotropic distribution of momentum should be considered.
\end{thebibliography}
\end{document}